\documentclass[journal=accacs,manuscript=article,layout=onecolumn]{achemso}
\usepackage[version=3]{mhchem} % Formula subscripts using \ce{}

\usepackage{adjustbox}
\usepackage{multirow}
\usepackage{mdframed}
\usepackage{rotating}
\usepackage{booktabs}
\usepackage{etoolbox}
\usepackage{enumitem}
\usepackage{subcaption}
\usepackage[nonumberlist]{glossaries}
\mciteErrorOnUnknownfalse
\usepackage{xcolor}
\usepackage{comment}
\usepackage{array}
\usepackage[hidelinks]{hyperref}
\usepackage[version=3]{mhchem}
\newcounter{magicrownumbers}
\preto\tabular{\setcounter{magicrownumbers}{0}}
\usepackage{todonotes}
\usepackage{longtable}

\newacronym{DFT}{DFT}{Density Functional Theory}
\newacronym{QE}{QE}{Quantum Espresso}
\newacronym{VASP}{VASP}{Vienna Ab initio Simulation Package}
\newacronym{S2EF}{\textit{S2EF}}{Structure to Energy and Forces}
\newacronym{IS2RE}{\textit{IS2RE}}{Initial Structure to Relaxed Energy}
\newacronym{IS2RS}{\textit{IS2RS}}{Initial Structure to Relaxed Structure}
\newacronym{OC20}{OC20}{Open Catalyst 2020}
\newacronym{ML}{ML}{Machine Learning}
\newacronym{TM}{TM}{Transition Metal}
\newacronym{FCC}{FCC}{Face-centered cubic}
\newacronym{MKM}{MKM}{Microkinetic Model}
\newacronym{MAE}{MAE}{Mean Absolute Error}

\glsaddall

\newcommand{\cmu}{Department of Chemical Engineering, Carnegie Mellon University}
\newcommand{\meta}{Fundamental AI Research (FAIR), Meta}

\title[]
  {CatTSunami: Accelerating Transition State Energy Calculations with Pre-trained Graph Neural Networks}

\author{Brook Wander}
\affiliation{\cmu}
\alsoaffiliation{\meta}

\author{Muhammed Shuaibi}
\affiliation{\meta}

\author{John R. Kitchin}
\affiliation{\cmu}

\author{Zachary W. Ulissi}
\affiliation{\meta}

\author{C. Lawrence Zitnick}
\affiliation{\meta}

\email{zitnick@meta.com}
\abbreviations{}

\let\oldmaketitle\maketitle
\let\maketitle\relax

\begin{document}



% \twocolumn[
% \begin{@twocolumnfalse}
\oldmaketitle

\begin{abstract}
 Direct access to transition state energies at low computational cost unlocks the possibility of accelerating catalyst discovery. We show that the top performing graph neural network potential trained on the OC20 dataset, a related but different task, is able to find transition states energetically similar (within 0.1 eV) to density functional theory (DFT)  $91\%$ of the time with a 28x speedup. This speaks to the generalizability of the models, having never been explicitly trained on reactions, the machine learned potential approximates the potential energy surface well enough to be performant for this auxiliary task. We introduce the Open Catalyst 2020 Nudged Elastic Band (OC20NEB) dataset, which is made of 932 DFT nudged elastic band calculations, to benchmark machine learned model performance on transition state energies. To demonstrate the efficacy of this approach, we replicated a well-known, large reaction network with 61 intermediates and 174 dissociation reactions at DFT resolution (40 meV). In this case of dense NEB enumeration, we realize even more computational cost savings and used just 12 GPU days of compute, where DFT would have taken 52 GPU years, a 1500x speedup. Similar searches for complete reaction networks could become routine using the approach presented here. Finally, we replicated an ammonia synthesis activity volcano and systematically found lower energy configurations of the transition states and intermediates on six stepped unary surfaces. This scalable approach offers a more complete treatment of configurational space to improve and accelerate catalyst discovery.
\end{abstract}
% \end{@twocolumnfalse}

%%%MAIN TEXT%%%%
% \hline
%%%%% Introduction
\section{Main}
% make it more clear that the sow is heterogeneous catalysis (slab model)
% Need to add some citations here
As global greenhouse gas emissions continue to raise the earth's temperature, there is an urgent need to cut emissions\cite{ipcc_report}. As we look to redesign our chemical and energy infrastructure to achieve a more sustainable future, while also seeking to meet burgeoning demands, there is a need to hasten the development of new processes to support our world's needs. Many of these processes will involve the reformation of energy and chemicals for which catalysts play the important role of improving efficiency. Methods to accelerate the discovery of catalysts will support the execution of our vision for a sustainable future. 

% If acceleration is achieved, it will unlock this value
Experimental catalyst discovery is a time consuming process of trial and error which relies on expert domain knowledge\cite{FARRUSSENG2008487, senkan2001combinatorial}. We envision a future where a step change in the acceleration of this critical process is realized using computational techniques. Today major gaps exist between what is treated computationally and what is realized experimentally. One source of disparity is the difficulty of accessing transition state energies, which are paramount for determining kinetic information\cite{xie2022achieving}. One common method used to calculate the energy of the transition state is the Nudged Elastic Band (NEB) method, which is very costly computationally when performed with sufficiently high accuracy Density Functional Theory (DFT)\cite{koistinen2019nudged,schlegel2011geometry}. For heterogeneous catalysts, there are many possible adsorbate and site configurations that could be used as input to the NEB calculation. Because of the high computational cost, very few of these possible configurations are sampled in practice, meaning there is little guarantee the correct barrier has been considered. We seek to accelerate the calculation of accurate NEBs using off-the-shelf, pretrained machine learned potentials (MLP), which unlocks three opportunities to improve the fidelity of computational studies: (1) acceleration and augmentation of reaction mechanism search, (2) acceleration and augmentation of kinetic studies, and (3) Use of kinetics directly in screening. All of these opportunities will serve to improve computational treatment of heterogeneous catalyst systems. In addition to these opportunities with acceleration of NEBs, the generalized nature of an MLP opens the possibility of applying it to NEB alternatives beyond the scope of this work such as the dimer method\cite{henkelman1999dimer}, growing string method\cite{jafari2017reliable}, and methods that rely on approximating the Hessian\cite{zeng2014unification, hermes2019accelerated, sharada2014finite}.

\textbf{Acceleration and augmentation of reaction mechanism search.} For reaction networks with significant complexity, it is challenging to elucidate which reaction pathways are important. Work has been done to automate the generation of possible reaction pathways\cite{rangarajan2012language, zeng2020reacnetgenerator, goldsmith2017automatic}. This leaves the job of figuring out, among the many possible pathways enumerated, which are of consequence to observed rates. To accomplish this, it is typical to rely on approximations for both thermochemical values and kinetic values\cite{ulissi2017address, liu2021reaction, kreitz2023automated}. Because these approximations only hold true when considering like sites or surfaces in heterogeneous catalysis, it greatly limits the broader applicability of the approach. This work can improve the approximations of kinetic values making identification of relevant pathways more accurate and also does not require assertions of surface or site type so it may be extended to varied materials as needed. 

\textbf{Acceleration and augmentation of kinetic studies}. It is common to use DFT studies to try to explain experimental outcomes to extract a deeper understanding of catalyst systems\cite{liu2021density, yang2022significance, hirunsit2015co2,  li2015heterogeneous}. These studies further our understanding of heterogeneous catalysts and how we may design them to be better. This work can accelerate such studies allowing the scope of studies to be expanded or increase the rate at which they are completed, thereby accelerating our understanding of heterogeneous catalysts.

\textbf{Use of kinetics directly in screening.} Computational approaches to screen catalyst materials rely on thermochemical descriptors which can be correlated with figures of merit like activity and selectivity\cite{greeley2006computational, takahashi2022synthesis, wander2022catlas}. It is, however, a grand outstanding challenge to screen catalyst candidates using kinetics directly\cite{bollini2023vision}. This would be transformative because it would allow for the reduction of assumptions and reliance on correlations which do not extrapolate well. As we will show, MLPs are able to perform NEB calculations at a fraction of the compute cost with reasonable accuracy. This unlocks the opportunity of using kinetics directly in screening.

% How people have tried to accelerate / approximate the TS energy in the past
There are two categories of approaches to circumnavigate the high computational cost of accessing transition state energies: approximating the energies and accelerating their calculation. The most foundational method of \textit{approximating} the transition state energy is using Br\o{}nsted-Evans-Polanyi (BEP) relations. It has been demonstrated, when considering similar active sites, that the activation energy may be linearly related to the reaction energy across surfaces\cite{logadottir2001bronsted, bligaard2004bronsted, munter2008bep}. This is the basis for correlating kinetic values of merit to thermochemical descriptors. Recently, a database of activation energies reported in literature has been compiled to perform Gradient Boosted Regression for the prediction of unknown activation energies\cite{hutton2023machine}. These approaches are very computationally inexpensive, but they lack extensibility, do not have great accuracies, and do not allow the transition state structure to be accessed.

For \textit{acceleration}, the general approach is to use a method that accesses energies and/or forces at a reduced computational cost to supplement or perform the necessary calculations. It is natural to turn to MLPs to accelerate NEB calculations because they can offer higher accuracy than traditional force fields while retaining a reduced computational cost when compared to sufficiently accurate DFT. Training an MLP to accelerate NEB calculations was first demonstrated for simple systems in 2016 by Peterson\cite{peterson2016acceleration}. Work has also been done to accelerate NEB calculations using Gaussian Process Regression\cite{koistinen2017nudged, koistinen2019nudged, torres2019low} and using on-the-fly active learning of a modified Behler-Parrinello neural network\cite{yang2021machine}. Recently, a generalized framework for training MLPs on the fly has been demonstrated for accessing energy barriers\cite{schaaf2023accurate} and for estimating the free energy of barriers, going beyond the harmonic approximation\cite{stocker2023estimating}. These prior contributions using machine learning to accelerate NEB calculations required the development of bespoke models for a specific system, which limits their applicability and acceleration potential. 

\begin{figure}[t]
    \centering
    \includegraphics[width=\textwidth]{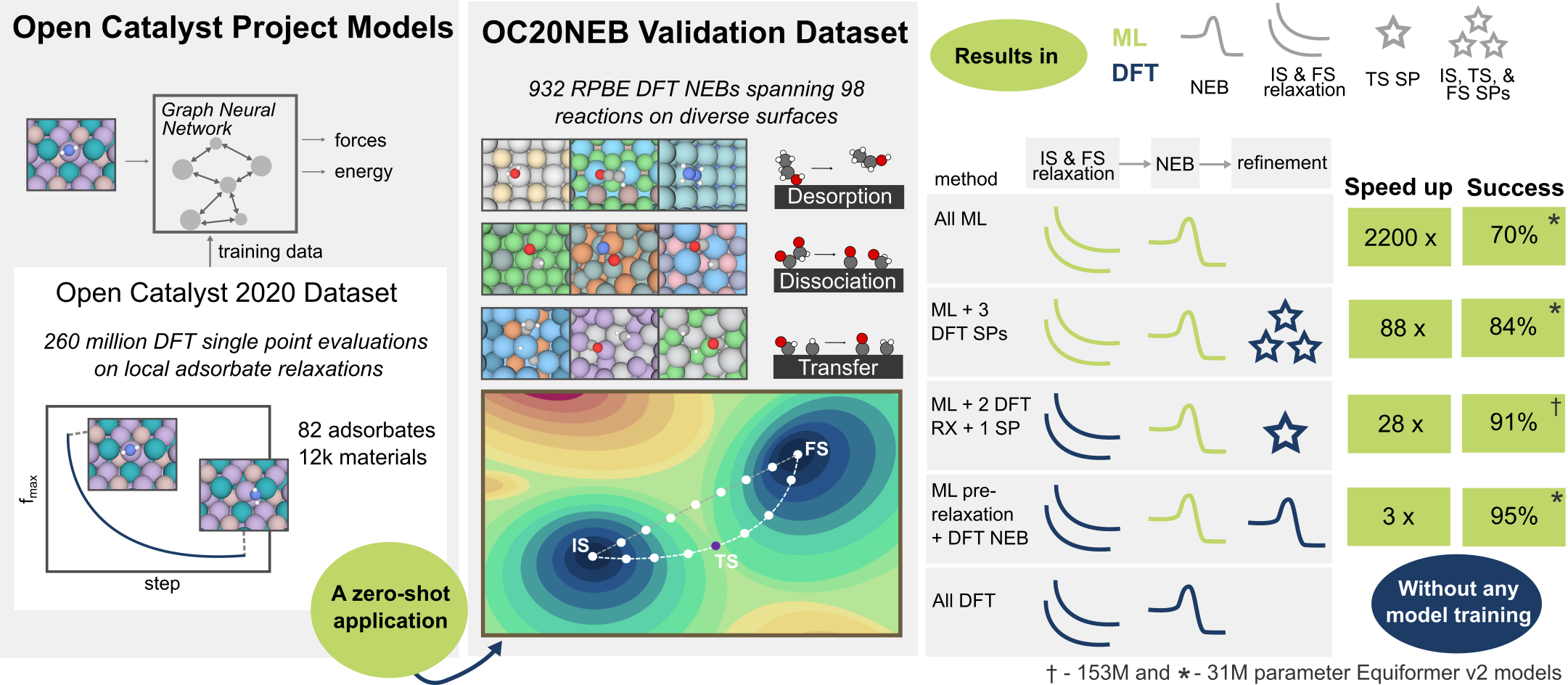}
    \caption{A summary of the work presented here. We demonstrate that models trained on local adsorbate relaxations perform well in a zero-shot application to NEB calculations. To do so we created a validation dataset of 932 RPBE NEB calculations. With this dataset we tested 4 ML approaches to accelerating NEBs. The results of this are shown in the left panel, where success is the proportion of systems where the ML accelerated approach gives an activation energy that agrees with the DFT calculated NEB within 0.1 eV. }
    \label{fig:summary}
\end{figure}

Here, we demonstrate that off-the-shelf models which have been pre-trained only on the OC20 dataset\cite{chanussot2021open} are able to perform NEB calculations that arrive at similar energy transition states 91\% of the time. The use of machine learning (ML) provides a 28x speed up across a variety of reactions on a diverse group of surfaces. The OC20 dataset consists of adsorbate relaxations only, a relevant but different task, making this a zero-shot application of the MLP. This speaks to the generalizability of the potential learned. To facilitate the assessment of model performance we present the OC20NEB dataset, a dataset composed of 932 revised Perdew-Burke-Ernzerhof (RPBE) DFT NEB calculations. A summary of these contributions may be seen in Figure \ref{fig:summary}.

We propose the CatTSunami framework to facilitate use of OC20\cite{chanussot2021open} pre-trained models to perform high-throughput ML-accelerated NEB calculations. To demonstrate the efficacy of CatTSunami, we apply it to two problems which address the potential to unlock value for two of the aforementioned opportunities. To address the opportunity to accelerate and augment reaction mechanism search, we searched for low energy transition states for 174 possible dissociation reactions for CO hydrogenation on the close-packed surface of rhodium. We compare our results to those presented by Ulissi et al.\cite{ulissi2017address} and show that our approach achieves DFT resolution (40 meV) at a low enough computational cost that all reactions could be considered explicitly. The full study, which considered 19,000 NEBs, would have taken 52 GPU years with DFT, but took just 12 GPU days with ML. This reveals the beneficial low-cost of ML-acceleration, which allows us to more exhaustively explore what previously relied on BEP scaling approximations and group additivity to estimate intermediate energies using a Gaussian Process model. To address the opportunity to accelerate and augment kinetic studies, we search for transition states in ammonia synthesis on the stepped surfaces of unary Ru, Fe, Pd, Co, Ni, and Rh to build a microkinetic volcano activity plot. We compare our results to those found by Vojvodic et al.\cite{vojvodic2014exploring} and find systematically lower energies. This again reveals the beneficial low computational cost of ML-accelerated NEBs, which allowed us to search more exhaustively for transition states to ultimately find lower energy configurations. Exhaustive search is particularly important for more complex catalyst surfaces, where it is difficult to rely on manual placement to find low energy configurations. The approach presented in this work is scalable and will drive faster insights into heterogeneous catalysts.

The contributions of this work are four-fold: (1) development of CatTSunami, a framework for high-throughput ML-accelerated NEBs, (2) curation of the OC20NEB dataset, consisting of 932 DFT NEB calculations, (3) assessment of baseline ML model performance on the dataset, and (4) demonstration of value added through two case studies.

% Lit review: https://docs.google.com/document/d/1m7y9BAvE5F0ZUAcaAMW7VWyJKZ09gJPRGKkIturkpNw/edit?usp=sharing 

%%%%% Results and Discussion
\section{Results}
\subsection{Dataset and Metrics}
\begin{figure}
    \centering
    \includegraphics[width=\textwidth]{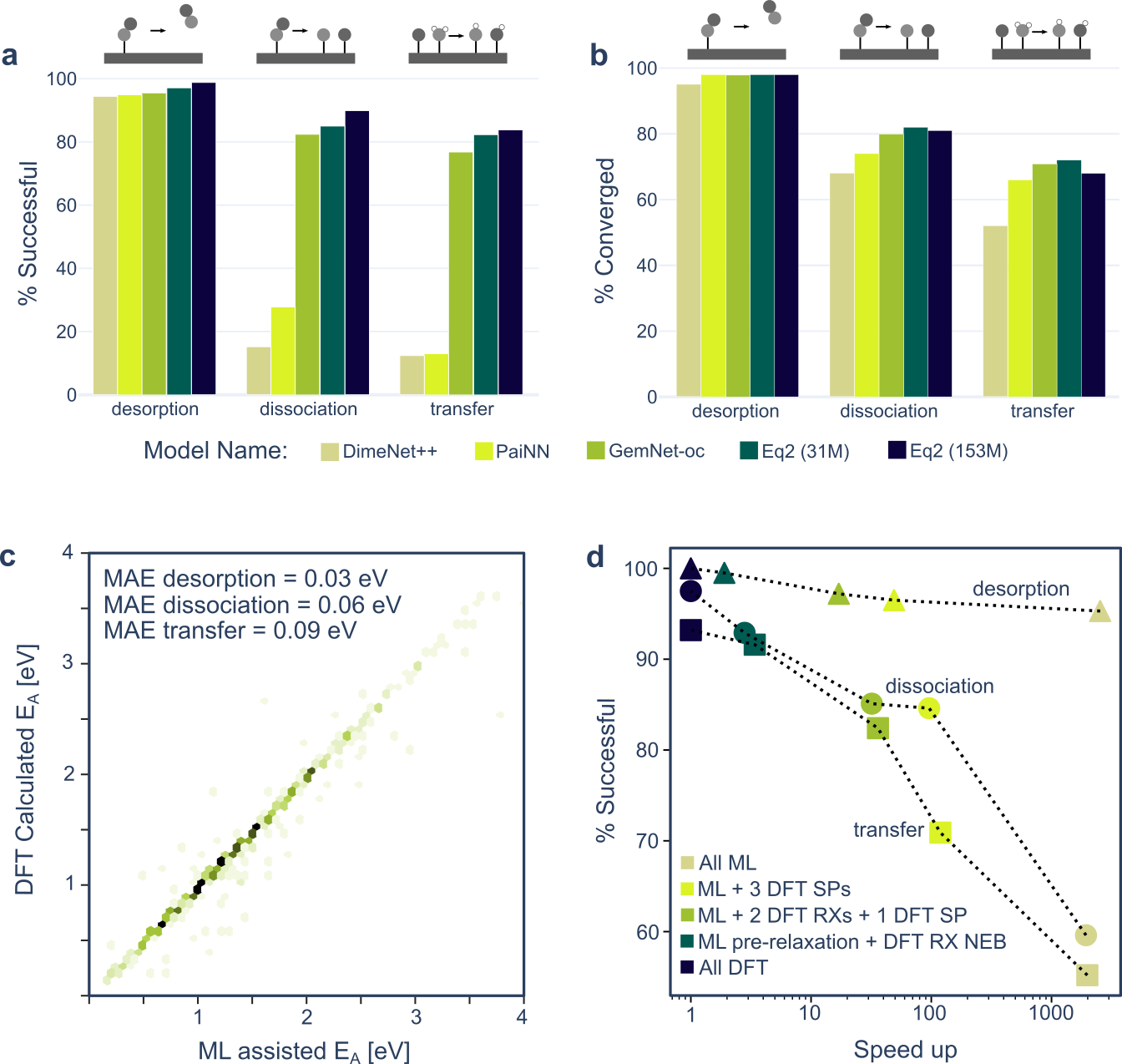}
    \caption{Baseline performance of five models according to the success metric. The percent of systems with an activation energy within 0.1 eV of the DFT determined activation is shown in (a) for systems where ML forces converged for the NEB. The percent of systems 
 which had converged forces is shown in (b). A parity plot of the DFT calculated activation energy versus the activation energy determined by an ML NEB with DFT relaxed initial and final frames, and a DFT single point on the transition state is shown in (c) for the Eq2 31M\cite{liao2023equiformerv2} model. The trade-off between computational speedup (as GPU seconds/ GPU seconds) and success is shown in (d) for four methods using the Eq2 31M\cite{liao2023equiformerv2} model}
    \label{fig:primary-results}
\end{figure}

The OC20NEB validation dataset consists of desorption, dissociation, and transfer reactions. Reactions were simulated on randomly generated surfaces containing up to three unique elements for substantial material diversity. It contains approximately 300 NEB calculations per reaction class, half of which were placed on materials that were reserved as out of domain and therefore did not appear in the OC20 training dataset. We have ensured that the initial and final frames are indicative of the reactant and product states intended (i.e. for a transfer reaction, over the reaction trajectory a transfer has in fact occurred). We did not, however, design an approach to determine if the transition state considered was of the type of interest. If the reaction coordinate contains multiple concerted steps, the step with the maximum transition state energy was considered. In our approach to enumerating systems, however, we limited the existence of multiple concerted steps by limiting the reaction path distance. The dataset is described in more detail in the Methods section.

To evaluate performance, we introduce a success rate as the proportion of ML calculations which find an energetically similar transition state.  We consider the ML NEB a success if the activation energy found is within 0.1 eV of the DFT activation energy.  We consider any transition state energy within 0.1 eV of the reactant or product state to be barrierless. For this case, we consider the ML NEB a success if both it and DFT find barrierless transition states. 

\subsection{Performance Assessment}
 We assess five pre-trained ML models ability to accelerate NEBs. The five models selected were: 153M and 31M parameter  Equiformer v2 models\cite{liao2023equiformerv2}, GemNet-OC\cite{gasteiger2022gemnet}, PaiNN\cite{schutt2021equivariant}, and DimeNet++\cite{gasteiger2020fast}. The pre-trained checkpoints for these models are available at https://github.com/FAIR-Chem/fairchem/tree/main/src/fairchem/core. The Equiformer v2 models were selected because of their state of the art performance on the OC20 dataset\cite{chanussot2021open} at the time of data generation. The others were selected to show the evolution of performance on this task as models have improved over time. We present the validation of five models using our success metric in Figure \ref{fig:primary-results}a. Models have been ordered by their performance on OC20 training metrics. We observe a positive correlation between the OC20 training metrics and performance on this auxiliary task. In general, desorptions have fewer consequential degrees of freedom than dissociations, which are simpler than transfers. This trend of complexity corresponds well with the performance metrics. We find highest success for desorptions (98.8\%), intermediate success for dissociations (89.9\%), and the lowest success for transfers (83.8\%) using the 153M parameter Equiformer v2 (Eq2 153M) model.

The DFT NEBs compared only include those which converged (i.e., those which achieved $f_{max} <$ 0.05 eV/\r{A} within the allowed number of steps). The proportions of ML NEB calculations which had converged forces ($f_{max} <$ 0.05 eV/\r{A}) where there exists a corresponding DFT NEB calculation which converged are shown in Figure \ref{fig:primary-results}b. We also observe a positive correlation between model performance on the OC20 training metrics and convergence. Models that more accurately predict forces and energies on the OC20 dataset, have a higher percent of calculations which converge. The one exception to this is the Equiformer v2 153 million parameter model which has a slightly lower convergence rate than less performant models for dissociations (81\% versus 82\%) and transfers (68\% versus 72\%). As is detailed in the Methods section, we used a fixed number of frames and a single spring constant for all NEB calculations. Convergence may be improved by adjusting these values or adopting an alternative optimization scheme.

Figure \ref{fig:primary-results}c shows a parity plot of the DFT calculated activation energy and the ML assisted activation energy using the 31M parameter Equiformer v2 (Eq2 31M) model\cite{liao2023equiformerv2}. The results for desorption, dissociation, and transfers are all shown, but the individual mean absolute errors (MAE) have been annotated. As we would expect from Figure \ref{fig:primary-results}a, desorptions have the lowest MAE (0.03 eV), dissociations have an intermediate MAE (0.06 eV), and transfers have the highest MAE (0.09 eV). More detailed parity plots for all of the models have been included in the Supplementary Information.

Figure \ref{fig:primary-results}d shows the trade-off between compute cost and success for the Eq2 31M model for four different ML accelerated approaches. To run a NEB, the initial and final frames must be relaxed. Subsequently, the intermediate frames must be iteratively optimized. At each optimization step, the forces and energies are evaluated. This provides various avenues for using an MLP to accelerate NEB calculations. The highest computational cost savings (2200x speed up) will be realized if ML is utilized for all force and energy evaluations \textbf{(All ML)}. This provides an average success of 70\%. The pre-trained MLPs have better accuracy for assessing forces than energies. A clear supplement to improve accuracy (84\% average) while achieving an 88x speedup is to use the MLP to perform all structure relaxations, but perform a DFT single point (SP) on the initial, transition state, and final structures \textbf{(ML + 3 DFT SPs)} to obtain more accurate energies. To perform a more direct comparison across models and between the DFT ground truth, we also considered the case where the initial and final frames were relaxed with DFT so the starting place for the NEB would be identical. Then the NEB was performed using the MLP and a single point was performed on the transition state \textbf{(ML + 2 DFT RXs + 1 DFT SP)}. This gives 88\% accuracy with a 28x speed up. Finally, we considered the case where the initial and final frames are relaxed with DFT, the NEB is pre-relaxed with ML, and subsequently relaxed with DFT \textbf{(ML pre-relaxation + DFT RX NEB)}, which will have the smallest computational cost savings (3x speed up) but provide DFT level accuracy. Now that the methods have been introduced, it is important to note that the results shown in Figure \ref{fig:primary-results}a-c use the method of ML + 2 DFT RXs + 1 DFT SP because, as mentioned, it allows for the most direct comparison of the ML NEB and the DFT NEB by removing the confounding variability introduced by having different initial and final states. A visual summary of the four acceleration methods is shown in the right side of Figure \ref{fig:summary}.

 The All DFT approach is not assigned 100\% success because in some cases a lower energy transition state along the band was found with DFT by the ML pre-relaxation + DFT RX NEB approach. Among the transfers, 7\% of the ML pre-optimized NEBs found a transition state more than 0.1 eV lower in energy than the pure DFT NEBs. Among the dissociations, 2\% were more than 0.1 eV lower in energy. In these cases, pre-optimizing with ML led to the discovery of a substantially lower energy transition state using DFT along the same NEB band. For desorptions and dissociations, there is a small change in performance between ML + 3 DFT SPs and ML + 2 DFT RX + 1 DFT SP, while this difference is large for transfers. Since these methods have different initial and final states, it must be that these differences have a more substantial impact on the transfer reactions, which is sensible given they have more consequential degrees of freedom. It is possible that some of the ML transition states in this case are valid, and even possibly lower in energy than the All DFT counterpart, but assessment of this was outside of the scope of this work. Tables of the results captured in Figure \ref{fig:primary-results} are included in the Supplementary Information.

\subsection{Comparison of Reaction Coordinate}
\begin{figure}
    \centering
    \includegraphics[width=\textwidth]{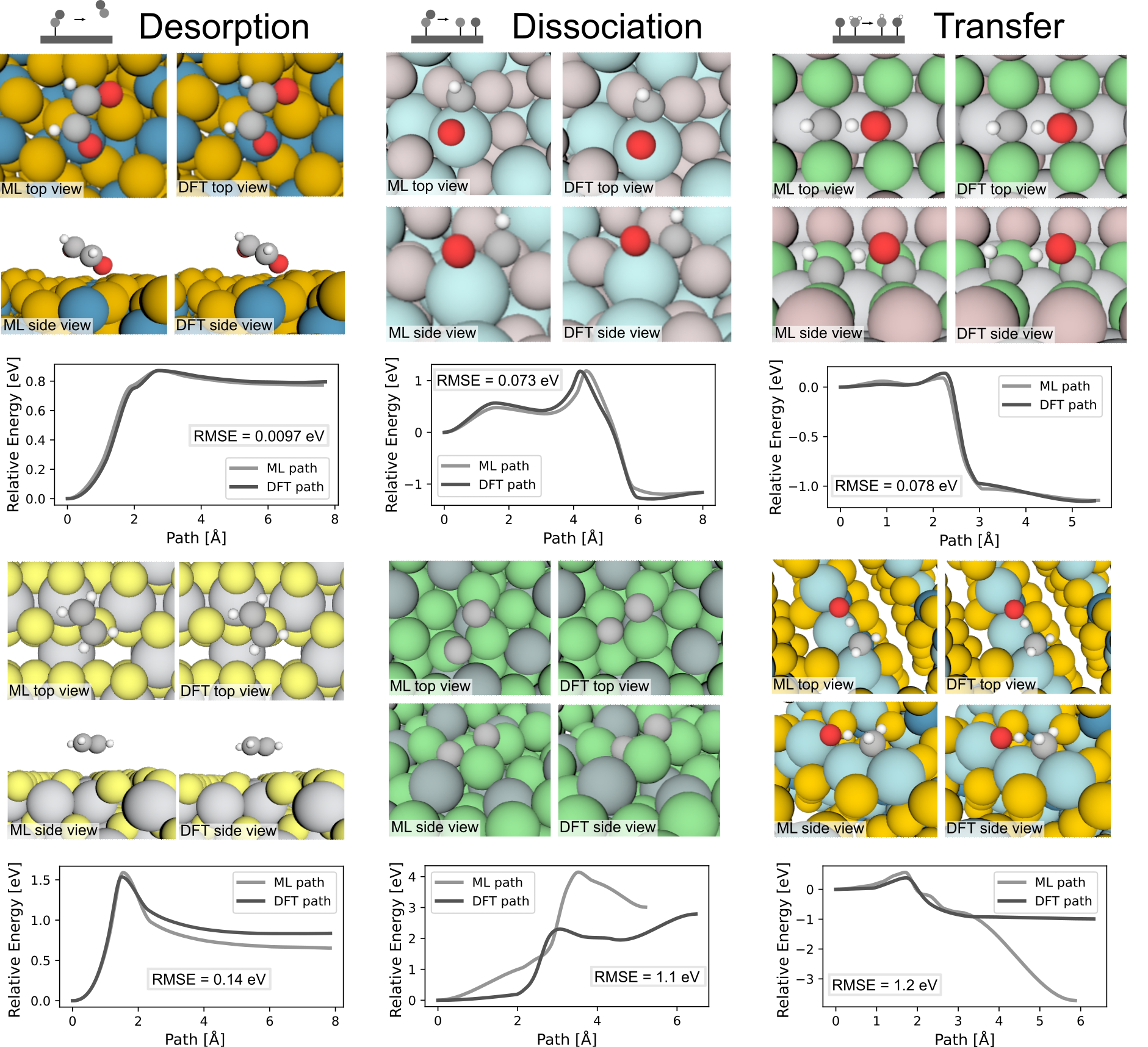}
    \caption{A comparison of the transition states and reaction coordinates found by ML and those found by DFT. The left column shows desorption, center shows dissociation, and left shows transfer. The top images compare the transition state structures for a randomly selected system, which correspond to the reaction coordinates below. The lower set have been selected among those with the highest root mean squared error (RMSE), integrated over the trajectory, to show a failure case.}
    \label{fig:ts-comparison}
\end{figure}

To assess the similarity of the reaction coordinates, we calculated the root mean squared error integrated over the reaction coordinate using Equation \ref{eq:1}. Figure \ref{fig:ts-comparison} shows a comparison of the DFT v. ML reaction coordinates and transition state structures for (top) randomly selected systems and (bottom) selected failure cases for each of the reaction types.  A histogram of RMSE of the Eq2 153M model\cite{liao2023equiformerv2} reaction coordinates for the full dataset is included in the Supplementary Information. It shows the vast majority of systems have very small RMSE, like the top examples in Figure \ref{fig:ts-comparison}. This speaks to the structural parity between ML and DFT determined transition states; going beyond the energetic comparison presented in Figure \ref{fig:primary-results}.

\subsection{Case Studies}
\begin{figure}
    \centering
    \includegraphics[width=\textwidth]{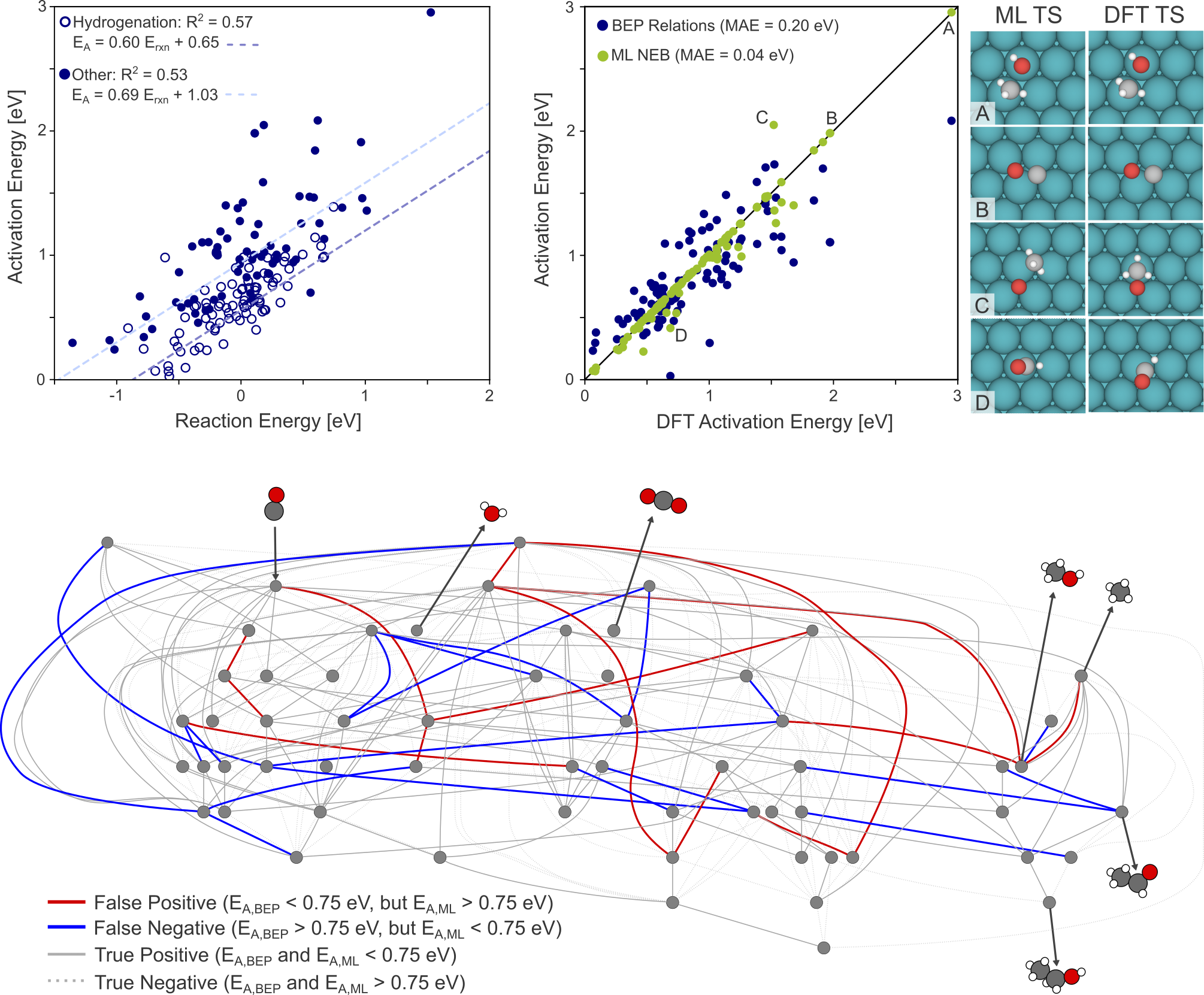}
    \caption{Top, left - BEP relations for dissociation reactions in the CO hydrogenation reaction network on the close-packed surface of Rh show significant scatter. Top, right - A parity plot comparing the performance of BEP relations and the ML + 3 DFT SP approach described in this work. To the right, images of the ML transition states and DFT transition states for four selected points are shown. Bottom - The CO hydrogenation reaction network. The nodes are reaction intermediates and the lines are reactions between them. The line color indicates the difference between the subset of reactions with activation energy less than 0.75 eV by BEP and by our ML accelerated approach.}
    \label{fig:case-study2}
\end{figure}

To demonstrate the efficacy of using pre-trained Open Catalyst Project (OCP) models to perform NEBs, we performed two case studies revisiting the works of Ulissi et al.\cite{ulissi2017address} and Vojvodic et al. \cite{vojvodic2014exploring}. The work of Ulissi et al.\cite{ulissi2017address} considers many reaction pathways for the hydrogenation of carbon monoxide on the rhodium close-packed surface to four important products including ethanol. There are approximately 200 reactions which connect the approximately 2000 pathways in the reaction network considered. To treat this problem, the authors approximated the energies of intermediate species using a Gaussian Process model and group additivity. They developed BEP relations to project the activation energy as a function of the reaction energy, which can be calculated using the energies of the reaction intermediates. For our work, we avoid approximations to the intermediate and activation energies by accelerating their acquisition using ML and the CatTSunami framework. Many local relaxations of reaction intermediates were performed with the Eq2 31M \cite{liao2023equiformerv2} model to find important low energy configurations on the surface. A search for low energy transition states was also performed for each of the 174 dissociation/ association reactions that appear in the network with the same model using our high-throughput workflow for enumerating NEBs with the ML + 3 DFT SPs approach. 

Using our automated approach, we were able to find transition states for 162 of the 174 reactions, without any adjustments to our standard parameters. Transition states were not found in the 12 cases because of (1) a lack of reactant configurations, (2) difficulty with convergence, or (3) removal at post processing because of insufficient frames or erroneous bond breaking/ forming events. These issues can be resolved with additional effort by increasing the number of frames, changing the spring constant, adopting and alternative optimization scheme, adjusting the automated enumeration parameters, or more exhaustively searching for reactant configurations where necessary.

Figure \ref{fig:case-study2} summarizes the results of explicitly calculating the dissociation/ association activation energies in the CO hydrogenation reaction network considered. The top, left plot shows the calculated activation energy as a function of the reaction energy for the reactions. In alignment with the prior work, we separated hydrogenations from other reaction classes. As can be seen, there is a poor correlation between these two values when considering a substantial diversity of reactions. This underscores the value of this method, which allows for the avoidance of reliance on BEP relations. To support this point, we calculated DFT NEBs on the set of reactions where we successfully found a transition state with ML.  A corresponding plot of the same values, but calculated with DFT rather than ML has been included in the SI. Of the reactions considered, 102 DFT NEBs converged using our standard number of frames and spring constant. Eight cases failed for reasons that could be resolved by rerunning. In Figure \ref{fig:case-study2}, the top, right plot shows a comparison of the performance of the ML + 3 DFT SP approach proposed here and BEP relations. In this case, the BEP relations are created using all 102 DFT NEBs, so it may be seen as an upper bound on the possible performance of BEP because it is the case where all of the activation energies are actually known. Using BEP relations results in a mean absolute error (MAE) of 0.20 eV. Using our approach with an 1500x speedup, results in an MAE of 0.04 eV, a marked improvement. For four select reactions, a comparison of the transition states found by ML and DFT are shown to the right. For A and B, which have ML and DFT activation energies in agreement, the structures appear identical. For C and D, which have ML and DFT activation energies in disagreement, the transition state structures are also very different. This brings into question whether the ML transition states are incorrect or just different from those found by DFT. 

The bottom portion of Figure \ref{fig:case-study2} shows the reaction network, where nodes are surface bound intermediate species and the lines are reactions. For visual simplicity, hydrogen has been omitted. The color explores the difference between the subset of reactions with activation energy less than 0.75 eV by BEP and by our ML accelerated approach. This is meant to explore the change in the network connectivity under a scenario where reactions with activation energies less than 0.75 eV are considered feasible. Red reaction would have been considered with BEP, but should not have according to the ML results. Blue reactions would not have been considered with BEP, but should have according to the ML results. The grey reactions agreed. Solid grey indicates that both ML and BEP arrive at an activation energy less than 0.75 eV. Dotted grey indicates that both ML and BEP arrive at an activation energy greater than 0.75 eV. We also compared our transition states to those used by Ulissi et al. to develop BEP relations. Despite the simplicity of the Rh (111) surface, in 4 cases (of 24), we believe we found lower energy transition states, but the prior work used the BEEF-vdW\cite{wellendorff2012density} functional, which makes direct energetic comparisons difficult. This is discussed and supported with a figure in the Supplementary Information.

\begin{figure}
    \centering
    \includegraphics[width=\textwidth]{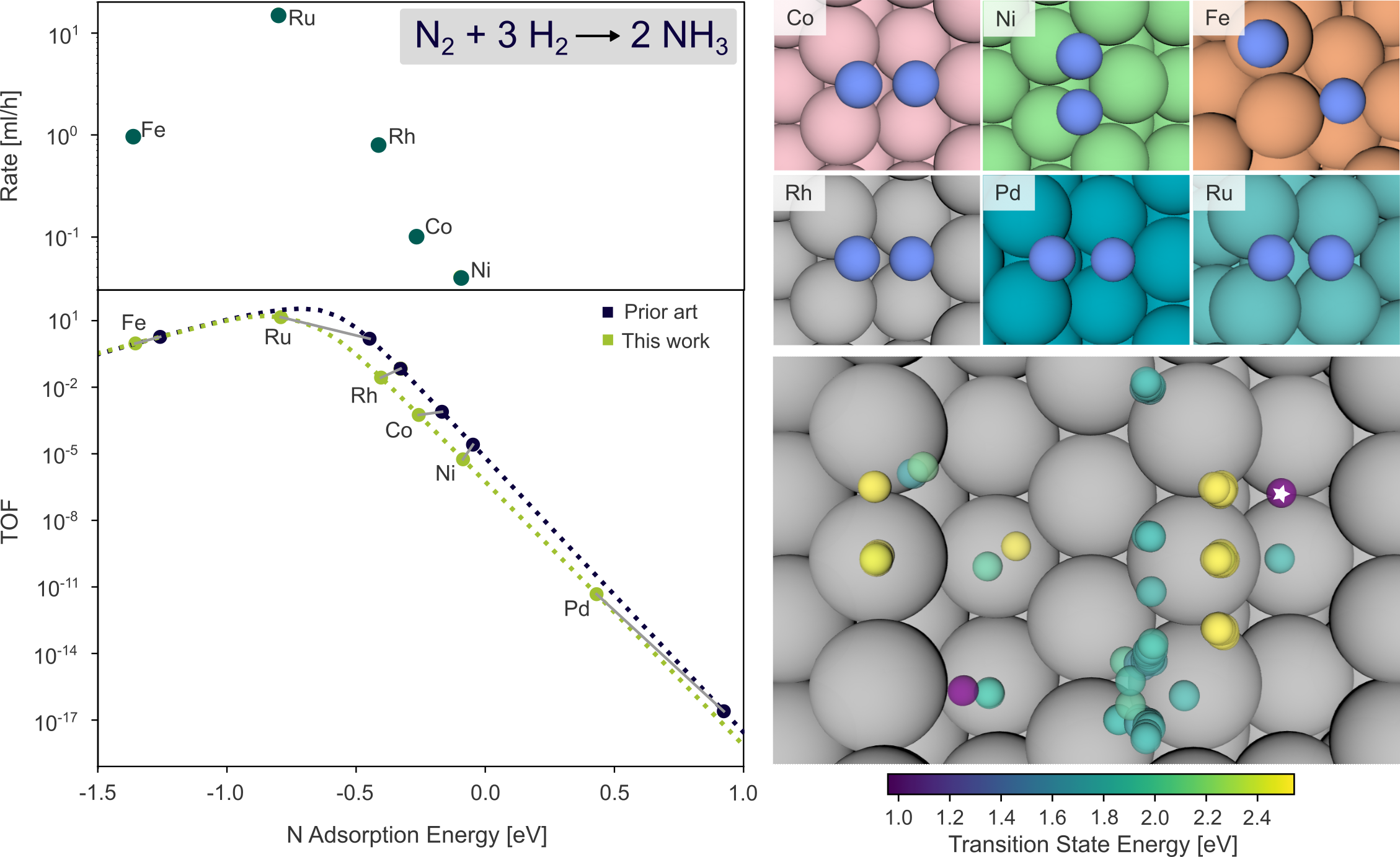}
    \caption{Left, top - experimental ammonia production rates using potassium promoted transition metal catalysts from Aiki et al.\cite{aika1973ammonia}. Left, bottom - a comparison of this work with that presented by Vojvodic et al. \cite{vojvodic2014exploring} on K-promoted stepped surfaces. Right, top - The N$_2$ transition state structures found in this work for each of the stepped unary surfaces. Right, bottom - The centers of mass of the various transition states considered in this work on the FCC Rh (211) surface. The centers of mass are colored according to the transition state energies and the lowest energy configuration is annotated with a star.}
    \label{fig:case-study1}
\end{figure}

The work of Vojvodic et al.\cite{vojvodic2014exploring} considers the development of an ammonia synthesis microkinetic model and volcano. This is used to explore the limits of the Haber-Bosch process and the scaling relations which make operation at high temperatures and high pressures a requirement today. For our work, we reproduce the activity volcano developed on the stepped surfaces of six unary metals. To do this, many local relaxations of reaction intermediates were performed with the Eq2 31M\cite{liao2023equiformerv2} model to find important low energy configurations on the surface. A search for low energy N$_2$ dissociation transition states was also performed with the same model using the CatTSunami framework with the ML + 3 DFT SPs approach. The models used in this work were not trained on data that includes spin polarization, so some additional refinement for Co and Ni was necessary. For more information, see the Methods.

Figure \ref{fig:case-study1} summarizes the results of reproducing an ammonia synthesis microkinetic model. The volcano curves for this work and the prior work are juxtaposed in the bottom left. Overall, there is good agreement between the two. For this work, the volcano is shifted to the left because lower energy configurations for reaction intermediates and transition states were found. For this work, ruthenium appears at the apex of the volcano and is more distinguished from iron and rhodium like we would expect given the experimental results in the top, left of Figure \ref{fig:case-study1}. Excluding the refinement to account for spin polarization, the work performed here could be completed in less than one day (10 GPU hours). If the same number of NEBs were performed using all DFT, it would have taken 1.7 GPU years. This could change the time it takes, and the very framework with which detailed kinetic studies are performed because transition state energies are so much easier to access by our approach.

The lowest energy transition states for each of the metals is shown in the top-right of Figure \ref{fig:case-study1}. For Co, Rh, Pd, Ru, and Fe we find that the lowest energy transition state occurs over the four-fold hollow site at the step. For Fe, this is a little obfuscated by significant adsorbate induced surface changes. For Ni, we find the lowest energy transition state at the three-fold hollow site atop the step. We did not consider the four-fold site in our automated enumeration. Still, the scaling relation between the transition state energy and the adsorption energy of monoatomic nitrogen was consistent with the other metals at the four-fold site (see Supplementary), so no further investigation was performed. It is likely that the transition state energy could have been lowered slightly by considering the four-fold site, but this would not have a large impact on the overall results. The lower, right figure shows the rhodium surface considered with centers of mass of every transition state colored by their energy. This illustrates the more exhaustive search enabled by our approach.

%%%%% Conclusions
\section{Conclusions}
In this work, we demonstrate that an off-the-shelf, zero-shot application of ML models is able to perform NEB calculations with an average of 91\% success and a 28x speed up, when compared to DFT. This speaks to the generalizability of the potential learned. To facilitate the assessment of model performance we present the OC20NEB dataset, a dataset composed of 932 revised Perdew-Burke-Ernzerhof (RPBE) Density Functional Theory (DFT) NEB calculations. In our assessment, we considered four different approaches to ML acceleration. None of these approaches require any model training, making inclusion into existing workflows facile. Use of our approach of NEB acceleration unlocks opportunities to improve and expand heterogeneous catalyst studies.

To ground this work, we perform two case studies which reproduce two prior works and directly explore the opportunity unlocked by this work. We find activation energies for 174 dissociation/ association reactions for the hydrogenation of carbon monoxide on the close-packed surface of rhodium. BEP relations achieve an MAE of 0.2 eV in the ideal scenario where all activation energies are known. ML accelerated NEBs achieve an MAE of 0.04 eV, closing this gap. This demonstrates the value of this work which makes calculating large numbers of transition states possible, while maintaining reasonable accuracy, where BEP has been used in the past.
We replicated a well-known, large reaction network with 61 intermediates and 174 dissociation reactions at DFT resolution with just 12 GPU days of compute. If this was performed with DFT, it would have taken 52 GPU years. Similar searches for complete reaction networks could be become routine using the approach presented here. With this, we are entering an era where complete reaction networks are enumerated and simulated quickly for complex surfaces. This will change the way we approach and build simplified reaction mechanisms. 

We also reproduce a microkinetic model of ammonia synthesis on stepped surfaces. We find good agreement overall between the volcano produced using ML acceleration and that reported in literature. We do, however, find systematically lower energy intermediates and transition states. It took less than a day of compute time to perform this study. This will change the way we approach performing detailed kinetic studies and use computational techniques to gain fundamental catalyst insights.

%%%%% Methods
\section{Methods}
\subsection{The CatTSunami framework}
We built tools to facilitate the generation of NEB calculations in a high-throughput fashion. These tools rely on the work presented by Lan et al.\cite{lan2023adsorbml}, which demonstrates the efficacy of OCP models for local relaxation and discovery of low energy sites. In alignment with the work of Lan et al.\cite{lan2023adsorbml}, the reactant(s) and product(s) are placed separately on the surface in many different configurations and local relaxations are performed using an ML model. The locally relaxed configurations are used to make initial and final frames for the NEB calculations with priority given to lower energy (by ML) configurations and proximity of adsorbates to one another and/ or to their initial position. The proposed final frames, are relaxed with an ML model and checks are performed to ensure the adsorbate connectivity is as expected. For more details about enumeration for the individual reaction types, see the Supplementary Information. 

From the initial and final frames, a reaction trajectory is interpolated across the desired number of frames. Iterative corrections are made to the linearly interpolated positions to avoid atomic overlap in a manner similar to IDPP\cite{smidstrup2014improved}. Ideally, the adsorbates should traverse the shortest path possible over the NEB. To do this, the absolute positions of the adsorbates in the initial and final frames are chosen so they are most proximate to one another. Care is taken to ensure issues interpolating over periodic boundary conditions would be avoided, which is important for making reasonable NEBs at high-throughput. Because the linear interpolation does not include the minimum image convention, the adsorbates are unwrapped to minimize euclidean distance over their bonds. This is done as a depth first search starting with the atom that is bound to the surface in the adsorbate. It is also enforced that the portions of the adsorbate(s) that maintain connectivity from the initial to the final frame move as a single moiety. This avoids the failure case where the minimum path for one atom in an adsorbate crosses a periodic boundary while another does not. Rather the minimum path of the binding atom is the path taken by the whole adsorbate. This still leaves the possibility of some poor NEB initializations, namely collisions across the trajectory, but the vast majority of initializations are reasonable. All the code that was used to generate NEB frames is available on GitHub: https://github.com/FAIR-Chem/fairchem/tree/main/src/fairchem/applications/cattsunami.

\subsection{The OC20NEB dataset}
The OC20NEB dataset consists of approximately 300 NEBs per reaction class. It was necessary to curate a database of reactions. This was done by hand for each of the reaction categories. The dissociation, transfer, and desorption databases contain 54, 25, and 19 reactions respectively. A full table of reactions has been provided in the Supplementary Information. Half of the NEBs were performed on materials that are reserved as out of domain (OOD) and the other half were on in domain (ID) materials in the OC20 dataset\cite{chanussot2021open}. There was not a significant difference in success between calculations performed on ID materials and OOD materials. Full results including these splits have been included in the Supplementary Information. For generation, a bulk was randomly selected from the desired domain among the 12k materials that appear in the OC20 dataset. These materials span 55 elements and contain up to 3 elements each. Although oxides have been excluded, other non-metals such as nitrides and sulfides are included along with interesting 2D materials. Next, a random Miller index, up to a maximum of 2, was selected. All distinct slabs for that miller index were enumerated, and one was randomly chosen.  Finally, a reaction was chosen at random to be combined with the slab. If the number of atoms in the first reactant-surface configuration exceeded 100 atoms, then the combination was discarded to avoid very computationally expensive calculations. The CatTSunami framework was used to initialize the NEB frames for DFT relaxation using the Eq2 31M\cite{liao2023equiformerv2} model for all reactant and product system initial relaxations. Ten frames were used for all NEBs.

\subsection{Running NEB calculations}
To run NEB calculations, we leveraged the existing precedent\cite{henkelman2000climbing, henkelman2000improved, kolsbjerg2016automated} provided by the Atomic Simulation Environment (ase)\cite{larsen2017atomic}, with a Broyden–Fletcher–Goldfarb–Shanno (BFGS) optimizer. To run ML NEBs in a more efficient manner we wrote a custom child class of the ase NEB class, which allows force evaluations to be made in a batched fashion. For all DFT NEBs and ML validation, calculation was considered converged if the maximum force perpendicular to the reaction coordinate was less than 0.05 eV/\r{A}. A climbing image\cite{henkelman2000climbing} scheme was imposed, where initially climbing image was turned off, but once the force was below 0.45 eV/\r{A} or 200 steps had been taken, climbing image was turned on. NEB relaxations were allowed to run for up to 300 steps. For all calculations, a fixed spring constant of 1 eV/\r{A}$^2$ was used.

\subsection{DFT calculation settings}
In alignment with the high-throughput nature of this work and required continuity with the training data, all DFT settings were algorithmically set in the same manner as the OC20 dataset\cite{chanussot2021open}. All calculations were performed with the Vienna Ab Initio Simulation Package (VASP)\cite{kresse1994ab, kresse1996efficiency, kresse1996efficient, kresse1999ultrasoft} with the revised Perdew-Burke-Ernzerhof (RPBE) functional\cite{perdew1996generalized, zhang1998comment}. For more details, please see the Supplementary Information.

\subsection{Reaction coordinate analysis}
To compare reaction coordinates between ML and DFT we employed Equation \ref{eq:1}, inspired by the work comparing equation of state models by Lejaeghere et al.\cite{lejaeghere2014error}. Here, $E(p)$ is the difference in the fitted energy profiles (between DFT and ML) over the reaction coordinate and P is the reaction path.

\begin{equation} \label{eq:1}
RMSE = \sqrt{\frac{\int_{0}^{P} E^2(p)dp}{P}}
\end{equation}

\subsection{Case study - reaction mechanism elucidation}
 We enumerated many NEBs to find low energy transition states. 14\% of the calculations that converged were otherwise incorrect. The two failure modes we observed were (1) the NEB had insufficient frames and did not sample the transition state and (2) there was an erroneous loss/gain of connectivity in bonds not intended to be broken or formed. We filtered out these calculations by implementing two checks (1) there exists a frame where the bond intended to be broken is elongated between 1.25x and 1.63x the initial bond distance and (2) ensuring the connectivity is the same throughout the NEB except the bond that is being broken. 

\subsection{Case study - exploring catalyst trends across materials}
For both the local relaxations of reaction intermediates and the NEBs, the Eq2 31M \cite{liao2023equiformerv2} model was used. For each of the lowest energy ML relaxed structures (i.e. the reaction intermediates and transition states) a DFT single point was performed to obtain the energy (ML + 3 DFT SPs). The same functional and referencing scheme was used in the prior work and this work, so energies should be directly comparable. One small difference between these works is the use of the stepped surface of the Im$\overline{3}$m bulk of Fe here rather than the P6$_3$/mmc, but these stepped surfaces are very similar.

In the original work, spin polarization was considered for Co and Ni, but it was assumed that the under reaction conditions, high coverages would lead to magnetic quenching on iron. In alignment with this we considered spin only for Co and Ni. Spin polarization is not included within the OC20 dataset\cite{chanussot2021open} and therefore is absent from the models. We performed single points with spin polarization for the Co and Ni systems and noticed that in some cases the residual forces were substantially higher than the other metals ($>$ 0.5 eV/\r{A}). This was the case for both the Co and Ni transition states and all Co intermediates. To treat this, we performed local relaxations for the Co intermediates with spin polarization turned on starting from the ML relaxed structures. We also performed dimer calculations on the Co and Ni transition states starting from the NEB determined TS. In this case where the model was ill suited to address our problem, it still allowed the identification of low energy configurations for further refinement.

In the original work, the calculations were repeated for ruthenium with potassium promotion. The difference in the energies with and without K-promotion were applied to the remaining metals. We applied these same corrections to energies so they could be more directly compared to the experimental results which include K-promotion. A volcano plot without K-promotion for this work and the prior work has been included in the SI. We also used the Gibbs energy corrections for the prior work for our work.

\begin{acknowledgement}
% I think there isnt one for funding? 
We would like to acknowledge Joseph Gauthier (TTU) and Andrew Medford (GT) for helpful correspondence on NEB foundational information and details of case study 1, respectively. We would also like to acknowledge Raffaele Cheula for advice on running dimer calculations for the case studies.
\end{acknowledgement}

\section{Contributions}
Brook Wander: Application of statistical, mathematical, computational, or other formal techniques to analyze or synthesize study data. Conducting a research and investigation process, specifically performing the experiments, or data/evidence collection. Development or design of methodology. Management activities to annotate (produce metadata), scrub data and maintain research data (including software code, where it is necessary for interpreting the data itself) for initial use and later re-use. Ideas; formulation or evolution of overarching research goals and aims. Programming, software development; implementation of the computer code and supporting algorithms; testing of existing code components. Preparation, creation and/or presentation of the published work, specifically visualization/data presentation. Preparation, creation and/or presentation of the published work, specifically writing the initial draft. Preparation, creation and/or presentation of the published work by those from the original research group, specifically critical review, commentary or revision – including pre- or post-publication stages. Part of this work was completed during an internship at Meta.

Muhammed Shuaibi: Programming, software development. Ideas; formulation or evolution of overarching research goals and aims. Preparation, creation and/or presentation of the published work by those from the original research group, specifically critical review, commentary or revision – including pre- or post-publication stages.

John R. Kitchin: Ideas; formulation or evolution of overarching research goals and aims. Preparation, creation and/or presentation of the published work by those from the original research group, specifically critical review, commentary or revision – including pre- or post-publication stages.

Zachary W. Ulissi: Ideas; formulation or evolution of overarching research goals and aims. Preparation, creation and/or presentation of the published work by those from the original research group, specifically critical review, commentary or revision – including pre- or post-publication stages.

C. Lawrence Zitnik: Development or design of methodology. Ideas; formulation or evolution of overarching research goals and aims. Preparation, creation and/or presentation of the published work by those from the original research group, specifically critical review, commentary or revision – including pre- or post-publication stages.
\clearpage
\bibliography{main}
\clearpage

\section{Supplementary Information}

\subsection{DFT settings}
DFT calculations were performed with the Vienna Ab Initio Simulation Package (VASP)\cite{kresse1994ab, kresse1996efficiency, kresse1996efficient, kresse1999ultrasoft} with periodic boundary conditions and the projector augmented wave (PAW) pseudopotentials \cite{kresse1999ultrasoft, blochl1994projector}. The external electrons were expanded in plane waves with kinetic energy cut-offs of 350 eV. Exchange and correlation effects were taken into account via the generalized gradient approximation and the revised Perdew-Burke-Ernzerhof (RPBE)\cite{perdew1996generalized, zhang1998comment} functional, because of its improved description of the energetics of atomic and molecular bonding to surfaces\cite{hammer1999improved}. Bulk and surface calculations were performed considering a K-point mesh for the Brillouin zone derived from the unit cell parameters as an on-the-spot method, employing the Monkhorst-Pack grid\cite{monkhorst1976special}.
\subsection{Reaction Network Elucidation}

\begin{figure}[H]
    \centering
    \includegraphics[width=0.9\textwidth]{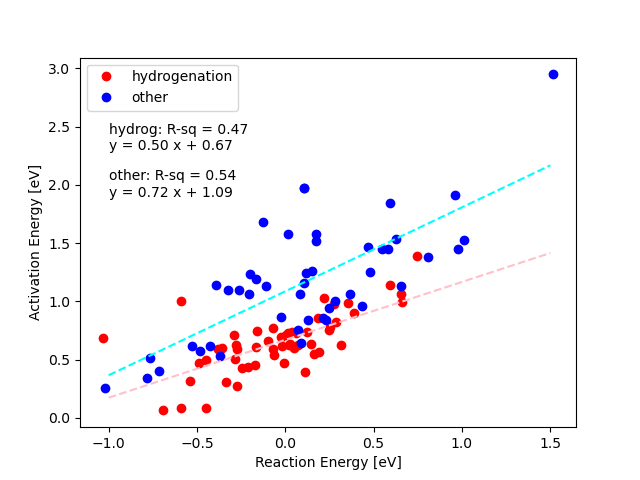}
    \caption{BEP relations for hydrogenations and other reactions across the CO hydrogenation reaction network.}
    \label{fig:BEP}
\end{figure}

\begin{figure}[H]
    \centering
    \includegraphics[width=0.9\textwidth]{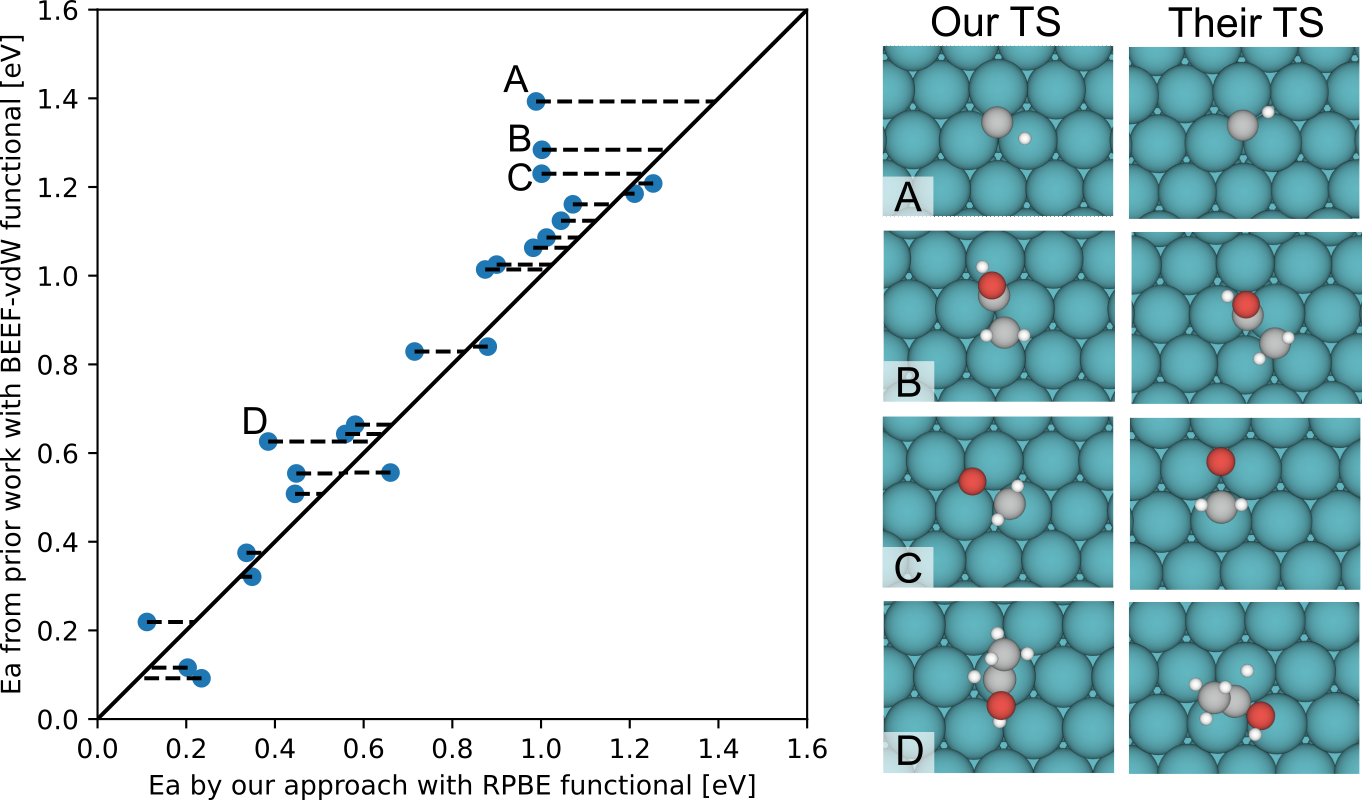}
    \caption{Left - A parity plot of the activation energy determined by our approach and the activation energy reported in the prior work\cite{ulissi2017address}. Right - images of the transition states for the 4 cases where the residual is greater than 0.2 eV}
    \label{fig:parity_ea}
\end{figure}

Figure \ref{fig:BEP} shows the BEP relation developed using DFT calculated NEBs. There were only 102 calculations that converged by our standard approach, which are pictured. Higher convergence could have been obtained by adjusting the spring constant, number of frames, and climbing image approach. Figure \ref{fig:parity_ea} - left shows the parity between the activation energy determined via our ML + 3 SPs approach and the values reported in the prior work\cite{ulissi2017address}. There is a functional difference, so we would expect there to be some scatter and making direct energetic comparisons is challenging. To the right are the 4 transition states which have a residual greater than 0.2 eV. In these cases, we believe that we have found lower energy transition states.

\subsection{Vibration calculations on DFT relaxed transition states}
\begin{table}[h]
\caption{Summary of the quantity of DFT calculations that were found to have imaginary vibrational modes.}
\label{tab:vib-modes}
\begin{tabular}{lrrr}
\toprule
 Reaction Type &  n Barrierless &  n Imaginary Frequency &  n No Imaginary Frequency \\
\midrule
 desorption &            249 &                     11 &                         0 \\
dissociation &             32 &                    215 &                         6 \\
transfer &             27 &                    174 &                         6 \\
\bottomrule
\end{tabular}
\end{table}
To check the validity of our force convergence criteria, we performed vibration calculations on a subset of our DFT NEB calculations. For cases where the reaction is not barrierless, and therefore the transition state is well defined, we find an imaginary vibrational mode the majority of the time as summarized in Table \ref{tab:vib-modes}. Here barrierless as defines as being within 0.1 eV of the reactant or product state.
\pagebreak
\subsection{Ammonia synthesis microkinetic model}
\begin{figure}[H]
    \centering
    \includegraphics[width=0.6\textwidth]{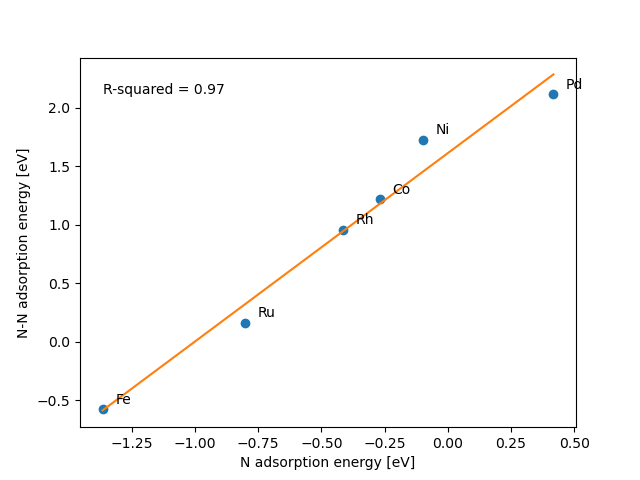}
    \caption{The scaling relation between N adsorption energy and N-N transition state energy as determined by our ML accelerated approach.}
    \label{fig:rmse_reaction_coord}
\end{figure}

\begin{figure}[H]
    \centering
    \includegraphics[width=0.6\textwidth]{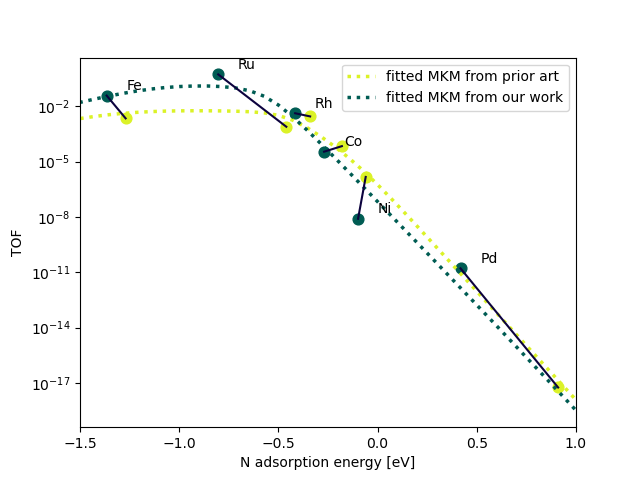}
    \caption{The activity volcano developed here and presented by Vojvodic et al.\cite{vojvodic2014exploring} without K-promotion. The points are the microkinetic rates using the raw energies, rather than the correlated activity.}
    \label{fig:volcano}
\end{figure}

Figure \ref{fig:volcano} shows activity volcano developed here and presented by Vojvodic et al.\cite{vojvodic2014exploring}. There are two key differences between the one presented here and that presented in the main paper (1) K-promotion has been excluded and (2) the points are the microkinetic rates using the raw energies, rather than the correlated activity. Because the rates are plotted explicitly, the effect of the slightly suboptimal Ni N-N transition state energy determined by this is noticeable. It is likely that if we would have calculated an energy at the four-fold hollow on the step this would be corrected. It can also be seen that in the prior work, Ru has a lower computationally determined rate, than Rh.
\subsection{RMSE in Reaction Coordinate Histograms}
\begin{figure}[h]
    \centering
    \includegraphics[width=\textwidth]{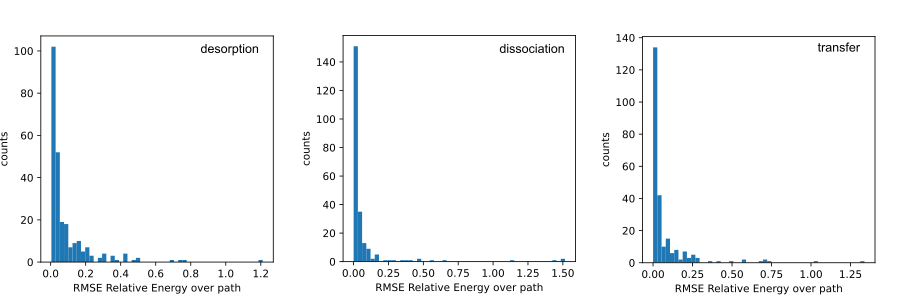}
    \caption{The distributions of observed RMSE between the DFT reaction coordinate at that which is obtained using ML.}
    \label{fig:rmse_reaction_coord_hist}
\end{figure}
\pagebreak

\subsection{Model Performance}
\subsubsection{All Models (ML + 2 DFT RX + 1 DFT SP)}
\begin{table*}[ht]
\small
\caption{Summary of model performance on the OC24NEB dataset. Here a calculation is successful if the activation energy falls within $\delta$ or is less than the DFT determined activation energy and has an imaginary vibrational mode on the transition state.}
\label{tab:results}
    \begin{tabular}{|l|l|l|l l|l l|}
    \hline
    \multirow{2}{*}{Reaction Type} &
    \multirow{2}{*}{Model Name} &
    \multirow{2}{*}{\% Converged} &
      \multicolumn{2}{c|}{\% within $\delta$ of converged} &
      \multicolumn{2}{c|}{\% within $\delta$ of all}\\
    &  &  &  $\delta$ = 0.1 eV  & $\delta$ = 0.05 eV &  $\delta$ = 0.1 eV  & $\delta$ = 0.05 eV  \\
\hline
desorption &  Eq2 (153M) &         98\% &                         98.8\% &                          98.8\% &                   96.8\% &                    96.4\% \\
dissociation &  Eq2 (153M) &         81\% &                         89.9\% &                          83.6\% &                   82.3\% &                    75.9\% \\
transfer &  Eq2 (153M) &         68\% &                         83.8\% &                          78.9\% &                   66.4\% &                    60.3\% \\
\hline
desorption &   Eq2 (31M) &         98\% &                         97.2\% &                          97.2\% &                   94.6\% &                    94.6\% \\
dissociation &   Eq2 (31M) &         82\% &                         85.1\% &                          73.2\% &                   76.5\% &                    64.8\% \\
transfer &   Eq2 (31M) &         72\% &                         82.4\% &                          70.2\% &                   67.6\% &                    55.6\% \\
\hline
desorption &   GemNet-oc &         98\% &                         95.6\% &                          94.4\% &                   94.1\% &                    92.9\% \\
dissociation &   GemNet-oc &         80\% &                         82.5\% &                          66.5\% &                   71.5\% &                    57.4\% \\
transfer &   GemNet-oc &         71\% &                         76.9\% &                          57.7\% &                   60.3\% &                    45.3\% \\
\hline
desorption &       PaiNN &         98\% &                         95.0\% &                          93.3\% &                   93.2\% &                    91.6\% \\
dissociation &       PaiNN &         74\% &                         27.9\% &                          14.2\% &                   24.0\% &                    11.8\% \\
transfer &       PaiNN &         66\% &                         13.1\% &                           6.3\% &                   10.9\% &                     5.3\% \\
\hline
desorption &   DimeNet++ &         95\% &                         94.4\% &                          93.2\% &                   89.5\% &                    88.4\% \\
dissociation &   DimeNet++ &         68\% &                         15.2\% &                          10.3\% &                   12.7\% &                     9.4\% \\
transfer &   DimeNet++ &         52\% &                         12.4\% &                           7.5\% &                    8.6\% &                     5.3\% \\
\hline
\end{tabular}
\end{table*}

\subsubsection{Single Model (ML + 3 DFT SPs)}
\begin{table*}[ht]
\small
\caption{Success assessment for an approach where only ML is used to relax the initial and final frames, then a NEB calculation is performed using the MLP, and finally DFT single points are performed on the initial, final, and transition state frames.}
\label{tab:ml_only}
\begin{tabular}{|l|l|l|l l|l l|}
    \hline
    \multirow{2}{*}{Reaction Type} &
    \multirow{2}{*}{Model Name} &
    \multirow{2}{*}{\% Converged} &
      \multicolumn{2}{c|}{\% within $\delta$ of converged} &
      \multicolumn{2}{c|}{\% within $\delta$ of all}\\
    &  &  &  $\delta$ = 0.1 eV  & $\delta$ = 0.05 eV &  $\delta$ = 0.1 eV  & $\delta$ = 0.05 eV  \\
\hline
desorption &  Eq2 (31M) &          98 &                         96.5\% &                          96.5\% &                   95.4\% &                    95.4\% \\
dissociation &  Eq2 (31M) &          83 &                         84.6\% &                          75.8\% &                   76.4\% &                    65.8\% \\
 transfer &  Eq2 (31M) &          71 &                         70.9\% &                          62.0\% &                   56.9\% &                    49.6\% \\
\hline
\end{tabular}
\end{table*}

\subsubsection{Single Model (ML pre-relaxation + DFT RX NEB)}
\begin{table}[H]
\small
\caption{Success assessment for an approach where first the initial and final frames are relaxed with DFT, then an NEB calculation is performed with the 31 million parameter Equiformer v2\cite{liao2023equiformerv2} model. Finally an NEB calculation is performed with DFT starting from the ML relaxed NEB}
\label{tab:dft-rx-from-ml}
\begin{tabular}{|l|l|l|l l|l l|}
    \hline
    \multirow{2}{*}{Reaction Type} &
    \multirow{2}{*}{Model Name} &
    \multirow{2}{*}{\% Converged} &
      \multicolumn{2}{c|}{\% within $\delta$ of converged} &
      \multicolumn{2}{c|}{\% within $\delta$ of all}\\
    &  &  &  $\delta$ = 0.1 eV  & $\delta$ = 0.05 eV &  $\delta$ = 0.1 eV  & $\delta$ = 0.05 eV  \\
\hline
desorption &  Eq2 (31M) &          98 &                         99.5\% &                          99.5\% &                   98.4\% &                    98.4\% \\
dissociation & Eq2 (31M) &          88 &                         92.9\% &                          91.0\% &                   85.7\% &                    84.1\% \\
transfer &  Eq2 (31M) &          79 &                         91.6\% &                          89.4\% &                   71.9\% &                    70.0\% \\
\hline
\end{tabular}
\end{table}

\subsubsection{Single Model (All ML)}
\begin{table*}[ht]
\small
\caption{Success assessment for the approach where the million parameter Equiformer v2\cite{liao2023equiformerv2} model is used for all force and energy calls.}
\label{tab:dft-rx-from-ml}
\begin{tabular}{|l|l|l|l l|l l|}
    \hline
    \multirow{2}{*}{Reaction Type} &
    \multirow{2}{*}{Model Name} &
    \multirow{2}{*}{\% Converged} &
      \multicolumn{2}{c|}{\% within $\delta$ of converged} &
      \multicolumn{2}{c|}{\% within $\delta$ of all}\\
    &  &  &  $\delta$ = 0.1 eV  & $\delta$ = 0.05 eV &  $\delta$ = 0.1 eV  & $\delta$ = 0.05 eV  \\
    \hline
desorption & Eq2 (31M) &          98 &                         95.3\% &                          93.7\% &                   94.2\% &                    92.7\% \\
dissociation &  Eq2 (31M) &          83 &                         59.6\% &                          38.3\% &                   52.7\% &                    33.9\%\\
transfer &  Eq2 (31M) &          71 &                         58.2\% &                          39.2\% &                   44.2\% &                    25.0\%  \\

\hline
\end{tabular}
\end{table*}

\subsubsection{Single Model - Speedup v. Success}
\begin{table*}[ht]
\small
\caption{The cost v. success tradeoff assessed for the 31 million parameter Equiformer v2 model}
\begin{tabular}{|p{0.25\linewidth}|ccc|ccc|}

\hline
   \multirow{2}{*}{Case} &
      \multicolumn{3}{c|}{\% Success} &
      \multicolumn{3}{c|}{Speed up}\\
    & desorption &  dissociation &  transfer &  desorption &  dissociation & transfer  \\
\hline
All ML &        95.3 &          59.6 &      55.2 &              2539.0 &                1936.0 &            1985.0 \\
\hline
ML + 3 DFT SPs &    96.5 &     84.6 &      70.9 &     49.0 &     96.0 &       119.0 \\
\hline
ML + 2 DFT RXs + 1 DFT SP &        97.2 &          85.1 &      82.4 &   17.0 &      32.0 &     36.0 \\
\hline
ML pre-relaxation + DFT RX NEB &  99.5 &   92.9 &   91.6 &      1.9 &         2.8 &   3.4 \\
\hline
All DFT &       100.0 &          97.5 &      93.2 &     1.0 &     1.0 &      1.0 \\
\hline
\end{tabular}
\end{table*}

\subsubsection{Model performance with split information (ML + 2 DFT RX + 3 DFT SPs)}
Roughly half of the NEB calculations performed were performed on materials which had been reserved for validation and are out of domain for the training set as defined in the OC20 dataset\cite{chanussot2021open}. The results segmented along these splits are shown in the table below. There is not an appreciable difference between calculations performed on in domain (ID) versus out of domain (OOD) materials. 

\begin{adjustbox}{center}
\tiny
% \caption={Model performance in finding a transition state within \delta of the DFT transition state or a lower energy transition state with an imaginary frequency. Results are shown here with in domain (ID) and out of domain (OOD) materials as defined in the OC20 dataset\cite{chanussot2021open}.}
\label{tab:split-results}
\begin{tabular}{|l|l|l|l l|l l| l| l l | l l |}
    \hline
    \multirow{2}{*}{Type} &
    \multirow{2}{*}{Model Name} &
    \multirow{2}{*}{ID Converged} &
      \multicolumn{2}{|c}{ID within $\delta$ converged} &
      \multicolumn{2}{|c|}{ID within $\delta$ all} &
      \multirow{2}{*}{OOD Converged} &
      \multicolumn{2}{|c}{OOD within $\delta$ converged} &
      \multicolumn{2}{|c|}{OOD within $\delta$ all}\\
    &  &  &  $\delta$=0.1 eV  & $\delta$=0.05 eV &  $\delta$=0.1 eV  & $\delta$=0.05 eV & &  $\delta$=0.1 eV  & $\delta$=0.05 eV &  $\delta$=0.1 eV  & $\delta$=0.05 eV  \\
\hline
desorption &  Eq2 (153M) &          98\% &    97.7\% &     97.7\% &    96.2\% &     95.5\% &           98\% &    100.0\% &     100.0\% &     97.5\% &      97.5\% \\
dissociation &  Eq2 (153M) &          80\% &    87.4\% &     83.9\% &    81.8\% &     77.3\% &           81\% &     92.2\% &      83.3\% &     82.7\% &      74.8\% \\
transfer &  Eq2 (153M) &          66\% &    87.9\% &     85.0\% &    68.3\% &     63.4\% &           69\% &     79.4\% &      72.2\% &     64.3\% &      56.6\% \\
desorption &   Eq2 (31M) &          99\% &    96.2\% &     96.2\% &    94.1\% &     94.1\% &           97\% &     98.3\% &      98.3\% &     95.1\% &      95.1\% \\
dissociation &   Eq2 (31M) &          81\% &    87.0\% &     78.7\% &    77.8\% &     68.9\% &           82\% &     83.5\% &      68.5\% &     75.3\% &      61.4\% \\
transfer &   Eq2 (31M) &          71\% &    83.2\% &     70.1\% &    66.0\% &     53.3\% &           73\% &     81.6\% &      70.4\% &     69.5\% &      58.2\% \\
 desorption &   GemNet-oc &         100\% &    93.9\% &     92.4\% &    93.9\% &     92.4\% &           97\% &     97.4\% &      96.6\% &     94.3\% &      93.4\% \\
dissociation &   GemNet-oc &          79\% &    89.0\% &     75.8\% &    75.2\% &     63.2\% &           81\% &     77.7\% &      59.5\% &     68.6\% &      52.9\% \\
transfer &   GemNet-oc &          73\% &    81.4\% &     63.6\% &    66.2\% &     50.3\% &           68\% &     71.7\% &      50.8\% &     53.9\% &      39.9\% \\
desorption &       PaiNN &          98\% &    92.2\% &     89.1\% &    90.8\% &     87.8\% &           98\% &     98.2\% &      98.2\% &     95.8\% &      95.8\% \\
dissociation &       PaiNN &          74\% &    30.1\% &     11.8\% &    24.2\% &      9.4\% &           74\% &     26.1\% &      16.2\% &     23.8\% &      13.9\% \\
transfer &       PaiNN &          66\% &    15.7\% &      6.3\% &    11.6\% &      5.1\% &           66\% &     10.0\% &       6.4\% &     10.2\% &       5.6\% \\
desorption &   DimeNet++ &          95\% &    93.7\% &     92.9\% &    89.0\% &     88.2\% &           94\% &     95.4\% &      93.5\% &     90.2\% &      88.5\% \\
dissociation &   DimeNet++ &          65\% &    18.2\% &     10.6\% &    13.4\% &      8.4\% &           71\% &     12.7\% &      10.1\% &     12.2\% &      10.1\% \\
transfer &   DimeNet++ &          59\% &    13.5\% &      7.3\% &     8.1\% &      5.1\% &           45\% &     10.8\% &       7.7\% &      9.1\% &       5.7\% \\
\hline

\end{tabular}
\end{adjustbox}
\pagebreak
\subsection{Desorption Reactions}
\begin{figure}[h]
    \centering
    \includegraphics[width=\textwidth]{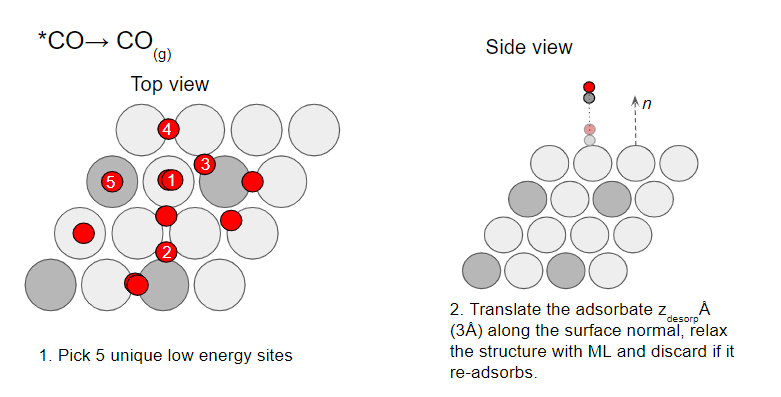}
    \caption{Details of desorption initial and final frame generation.}
    \label{fig:desorption-frame-gen}
\end{figure}
Starting from many ML relaxed configurations of the reactant adsorbed to the surface, the five lowest energy configurations of the reactant were selected NEB frames. Building the proposed final frame is shown in Figure \ref{fig:desorption-frame-gen}. The adsorbate is simply translated 3\r{A} along the surface normal for each of the product frames. The translated adsorbates are then allowed to relax using an MLP. Any frames where the product re-adsorbs or some other anomaly occurs were discarded. Among the remaining final frames, one is randomly selected to be interpolated with the reactant frame to run a DFT NEB. This process is facilitated with the  \texttt{AutoFrameDesorption} class. All reactions used for dataset generation are shown in the table below.
{\footnotesize
\begin{longtable}{ll}
\toprule
index &                  reaction  \\
\midrule
0  &              *CO $\rightarrow$ CO$_(g)$ \\
1  &              *N$_2$ $\rightarrow$ N$_2$$_(g)$  \\
2  &            *NH$_3$ $\rightarrow$ NH$_3$$_(g)$  \\
3  &            *OH$_2$ $\rightarrow$ H$_2$O$_(g)$  \\
4  &            *CH$_4$ $\rightarrow$ CH$_4$$_(g)$  \\
5  &        CH$_2$*CO $\rightarrow$ CH$_2$CO$_(g)$  \\
6  &     *CHO*CHO $\rightarrow$ CHOCHO$_(g)$  \\
7  &     *COH*COH $\rightarrow$ COHCOH$_(g)$  \\
8  &     *CH$_2$*CH$_2$ $\rightarrow$ CH$_2$CH$_2$$_(g)$ \\
9  &         *CH*CH $\rightarrow$ CHCH$_(g)$  \\
10 &    *CHOHCH$_2$ $\rightarrow$ CHOHCH$_2$$_(g)$  \\
11 &          *NHNH $\rightarrow$ NHNH$_(g)$  \\
12 &      *NH$_2$NH$_2$ $\rightarrow$ NH$_2$NH$_2$$_(g)$ \\
13 &              *NO $\rightarrow$ NO$_(g)$  \\
14 &        *OHNH$_2$ $\rightarrow$ HONH$_2$$_(g)$  \\
15 &          *ONOH $\rightarrow$ ONOH$_(g)$  \\
16 &  *OHCH$_2$CH$_3$ $\rightarrow$ CH$_3$CH$_2$OH$_(g)$  \\
17 &      *OCHCH$_3$ $\rightarrow$ CH$_3$CHO$_(g)$  \\
18 &        *OHCH$_3$ $\rightarrow$ CH$_3$OH$_(g)$  \\
\bottomrule
\end{longtable}}
\pagebreak
\subsection{Dissociation Reactions}
\begin{figure}[h]
    \centering
    \includegraphics[width=\textwidth]{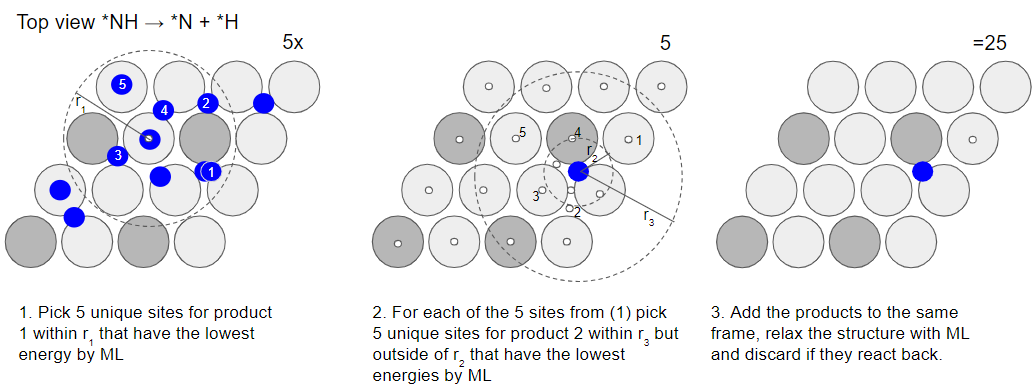}
    \caption{Details of dissociation initial and final frame generation.}
    \label{fig:dissociation-frame-gen}
\end{figure}

Starting from many ML relaxed configurations of the reactant and products individually placed on the surface, the lowest energy configuration of the reactant was selected to propose NEB frames. Building the proposed final frame is shown in Figure \ref{fig:dissociation-frame-gen}. Product 1 is defined as the product with the same binding atom as the reactant. First, five placements for product 1 were selected as the 5 lowest energy (by ML) configurations within $r_1 = 2$\r{A}. Next, for each of the 5 placements of product 1, we consider 5 proximate placements for product 2, giving a total of 25 final frames. Similarly, the 5 lowest energy configurations are chosen, within a shell of inner and outer radii of $r_2 = 1$\r{A} and $r_3 = 3$\r{A}, respectively. Then the adsorbate atoms of product 1 and product 2 are concatenated into the same frame and relaxed with ML. Any frames where the products react or some other anomaly occurs were discarded. Among the remaining final frames, one is randomly selected to be interpolated with the reactant frame to run a DFT NEB. This process is facilitated with the  \texttt{AutoFrameDissociation} class. All reactions used for dataset generation are shown in the table below.
{\footnotesize
\begin{longtable}{ll}
%\caption{All dissociation reactions in the curated database, which appear in the OC23-NEB dataset.}
%\label{tab:dissociation-rxns}
\toprule
index &                  reaction  \\
\midrule
0  &           *OH $\rightarrow$ *O + *H  \\
1  &           *CO $\rightarrow$ *C + *O  \\
2  &           *CH $\rightarrow$ *C + *H  \\
3  &         *OOH $\rightarrow$ *O + *OH  \\
4  &         *CH$_2$ $\rightarrow$ *CH + *H  \\
5  &         *CHO $\rightarrow$ *CO + *H  \\
6  &         *CHO $\rightarrow$ *CH + *O  \\
7  &         *COH $\rightarrow$ *C + *OH \\
8  &        *CH$_3$ $\rightarrow$ *CH$_2$ + *H  \\
9  &      *CH$_2$*O $\rightarrow$ *CH$_2$ + *O  \\
10 &      *CH2*O $\rightarrow$ *CHO + *H  \\
11 &       *CHOH $\rightarrow$ *CHO + *H \\
12 &       *CHOH $\rightarrow$ *COH + *H \\
13 &       *CHOH $\rightarrow$ *CH + *OH  \\
14 &        *CH$_4$ $\rightarrow$ *CH$_3$ + *H  \\
15 &     *OHCH$_3$ $\rightarrow$ *OH + *CH$_3$ \\
16 &     *OHCH$_3$ $\rightarrow$ *OCH$_3$ + *H  \\
17 &          *C*C $\rightarrow$ *C + *C  \\
18 &         *CCO $\rightarrow$ *C + *CO  \\
19 &         *CCH $\rightarrow$ *C + *CH \\
20 &       *CHCO $\rightarrow$ *CH + *CO \\
21 &       *CHCO $\rightarrow$ *CCO + *H \\
22 &       *CHCO $\rightarrow$ *CCH + *O \\
23 &       *CCHO $\rightarrow$ *C + *CHO  \\
24 &       *CCHO $\rightarrow$ *CCH + *O  \\
25 &     *COCHO $\rightarrow$ *CHCO + *O  \\
26 &     *COCHO $\rightarrow$ *CO + *CHO \\
27 &     *CCHOH $\rightarrow$ *C + *CHOH  \\
28 &     *CCHOH $\rightarrow$ *CCH + *OH  \\
29 &       *CCH$_2$ $\rightarrow$ *C + *CH$_2$  \\
30 &       *CCH$_2$ $\rightarrow$ *CCH + *H  \\
31 &      *CH*CH $\rightarrow$ *CH + *CH  \\
32 &     CH$_2$*CO $\rightarrow$ *CO + *CH2  \\
33 &     CH$_2$*CO $\rightarrow$ *CCH$_2$ + *O  \\
34 &     *CHCHO $\rightarrow$ *CH + *CHO  \\
35 &     CH*COH $\rightarrow$ *CHCO + *H  \\
36 &     CH*COH $\rightarrow$ *COH + *CH  \\
37 &  *COCH$_2$O $\rightarrow$ *CO + *CH$_2$*O  \\
38 &  *CHO*CHO $\rightarrow$ *CHO + *CHO  \\
39 &   *COHCHO $\rightarrow$ *COH + *CHO  \\
40 &   *COHCOH $\rightarrow$ *COH + *COH  \\
41 &       *CCH$_3$ $\rightarrow$ *C + *CH$_3$  \\
42 &     *CHCH$_2$ $\rightarrow$ *CH + *CH$_2$  \\
43 &     *COCH$_3$ $\rightarrow$ *CO + *CH$_3$  \\
44 &   *OCHCH$_2$ $\rightarrow$ *CHO + *CH$_2$  \\
45 &           *N$_2$ $\rightarrow$ *N + *N  \\
46 &           *NH $\rightarrow$ *N + *H  \\
47 &           *NO $\rightarrow$ *N + *O  \\
48 &           *CN $\rightarrow$ *C + *N  \\
49 &       *NONH $\rightarrow$ *NO + *NH  \\
50 &         *NH$_2$ $\rightarrow$ *NH + *H  \\
51 &        *NH$_3$ $\rightarrow$ *NH$_2$ + *H  \\
52 &     CO*COH $\rightarrow$ *COH + *CO \\
53 &           *H$_2$ $\rightarrow$ *H + *H  \\
\bottomrule
\end{longtable}}
\pagebreak
\subsection{Transfer Reactions}
\begin{figure}[h]
    \centering
    \includegraphics[width=\textwidth]{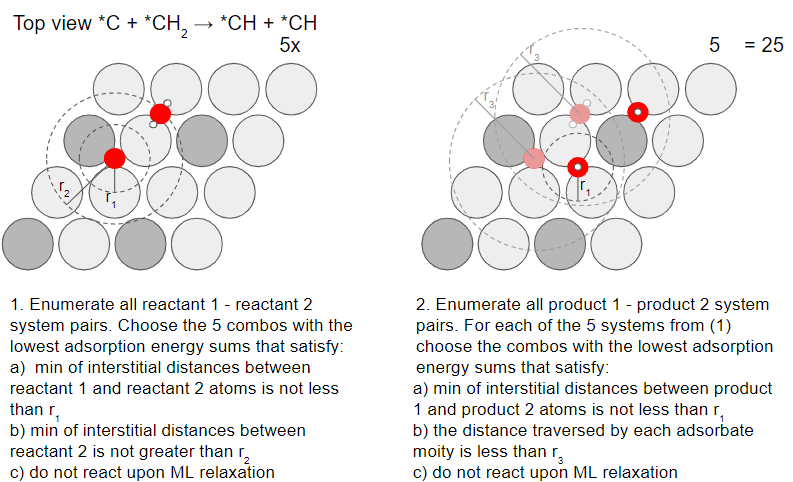}
    \caption{Details of dissociation initial and final frame generation.}
    \label{fig:transfer-frame-gen}
\end{figure}

Starting from many ML relaxed configurations of the reactants and products individually placed on the surface, all combinations of reactant 1 and reactant 2 are considered. Building the proposed final frame is shown in Figure \ref{fig:transfer-frame-gen}. The 5 reactant 1 - reactant 2 combinations with the lowest pseudo-adsorption energy (the sum of the individual reactant adsorption energies) which satisfy the following criteria are selected: (1) The minimum interstitial distance between any atom in reactant 1 and any atom in reactant two is no less than $r_1$ = 1 \r{A}, (2) The minimum interstitial distance between any atom in reactant 1 and any atom in reactant 1 is no greater than $r_2$ = 2 \r{A}, (3) upon relaxing the concatenated reactant 1 - reactant 2 system with ML, it does not react or have other anomalous behavior.  Next, for each of the reactant combinations, the 5 product 1 - product 2 combinations with the lowest pseudo-adsorption energy (the sum of the individual product adsorption energies) which satisfy the following criteria are selected: (1) The minimum interstitial distance between any atom in product 1 and any atom in product 2 is no less than $r_1$ = 1 \r{A}, (2) The distance traversed by each of the adsorbate moities is no greater than $r_3$ = 3 \r{A}, (3) upon relaxing the concatenated product 1 - product 2 system with ML, it does not react or have other anomalous behavior. Among the frame sets, one is randomly selected to be interpolated with the reactant frame to run a DFT NEB. This process is facilitated with the  \texttt{AutoFrameTransfer} class. All reactions used for dataset generation are shown in the table below.
{\footnotesize
\begin{longtable}{ll}
\toprule
index &   reaction  \\
\midrule
0  &            *OH + *CH$_2$ $\rightarrow$ *O + *CH$_3$  \\
1  &             *OH + *CH $\rightarrow$ *O + *CH$_2$ \\
2  &             *C + *CH$_2$ $\rightarrow$ *CH + *CH  \\
3  &          *CH$_3$ + *CH $\rightarrow$ *CH$_2$ + *CH$_2$ \\
4  &           *OH$_2$ + *CH $\rightarrow$ *OH + *CH$_2$ \\
5  &           *CHO + *CH $\rightarrow$ *CO + *CH$_2$ \\
6  &           *COH + *CH $\rightarrow$ *CO + *CH$_2$  \\
7  &             *N + *CH$_2$ $\rightarrow$ *NH + *CH \\
8  &   *OCH2CH$_3$ + *CH $\rightarrow$ *OCHCH$_3$ + *CH$_2$  \\
9  &           *NH + *CH$_2$ $\rightarrow$ *NH$_2$ + *CH  \\
10 &  *CH$_2$CH$_2$OH + *C $\rightarrow$ *CH$_2$*CH$_2$ + *COH \\
11 &     *NH$_2$NH$_2$ + *NO $\rightarrow$ *NH$_2$ + *ONNH$_2$ \\
12 &           *NO + *CH$_2$ $\rightarrow$ *ONH + *CH  \\
13 &     *COCH$_3$ + *COH $\rightarrow$ *CO + *COHCH$_3$ \\
14 &  *CH$_2$*CH$_2$ + *CH$_3$ $\rightarrow$ *CH$_2$CH$_3$ + *CH$_2$  \\
15 &       *CH$_2$*O + *C*C $\rightarrow$ *CH$_2$ + *CCO  \\
16 &    *CHO*CHO + *CO $\rightarrow$ *CHO + *COCHO  \\
17 &     *NO$_2$NO$_2$ + *NO $\rightarrow$ *NO$_2$ + *ONNO$_2$  \\
18 &         *C + *OHCH$_3$ $\rightarrow$ *CCH$_3$ + *OH  \\
19 &          *N + *NHNH $\rightarrow$ *N*NH + *NH  \\
20 &         *CH2OH + *H $\rightarrow$ *CH$_2$ + *OH$_2$ \\
21 &               *H + *C*C $\rightarrow$ *CH+ *C  \\
22 &            *CH$_4$ + *C $\rightarrow$ *CH$_3$ + *CH  \\
23 &          *CCH$_3$ + *C $\rightarrow$ *CCH$_2$ + *CH  \\
24 &             *OOH + *C $\rightarrow$ *O + *COH  \\
\bottomrule
\end{longtable}}
\pagebreak
\subsection{Parity and Residual Plots}

\begin{figure}[H]
    \centering
    \includegraphics[width=0.9\textwidth]{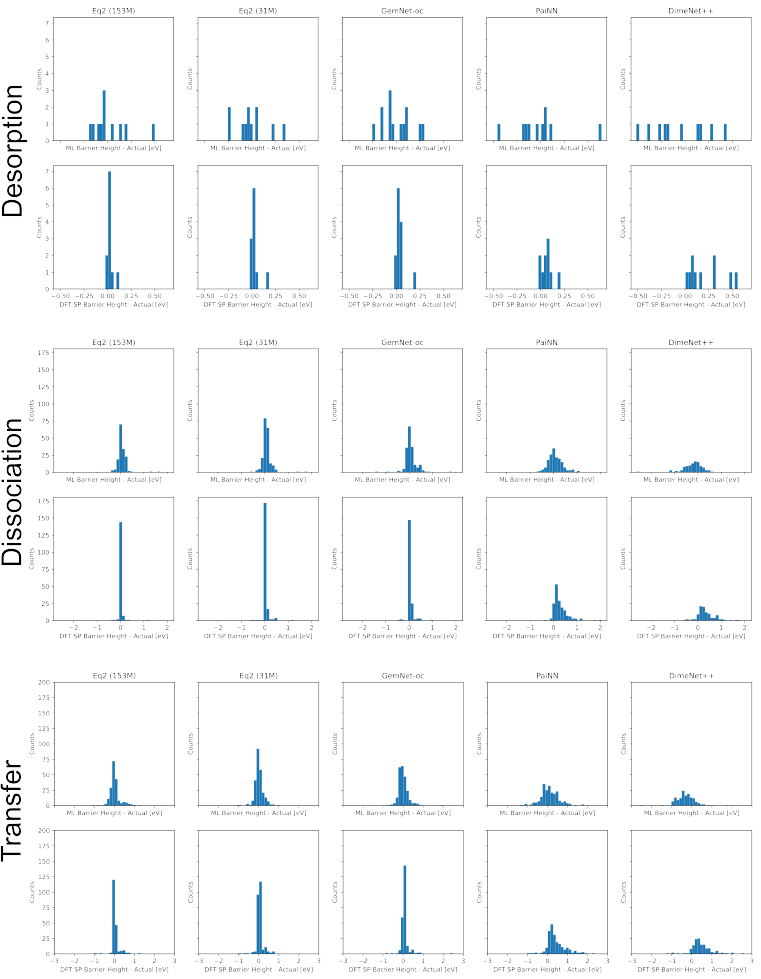}
    \caption{Residual distributions for all models across the three reaction types}
    \label{fig:residuals}
\end{figure}

\begin{figure}[ht]
    \centering
    \includegraphics[width=\textwidth]{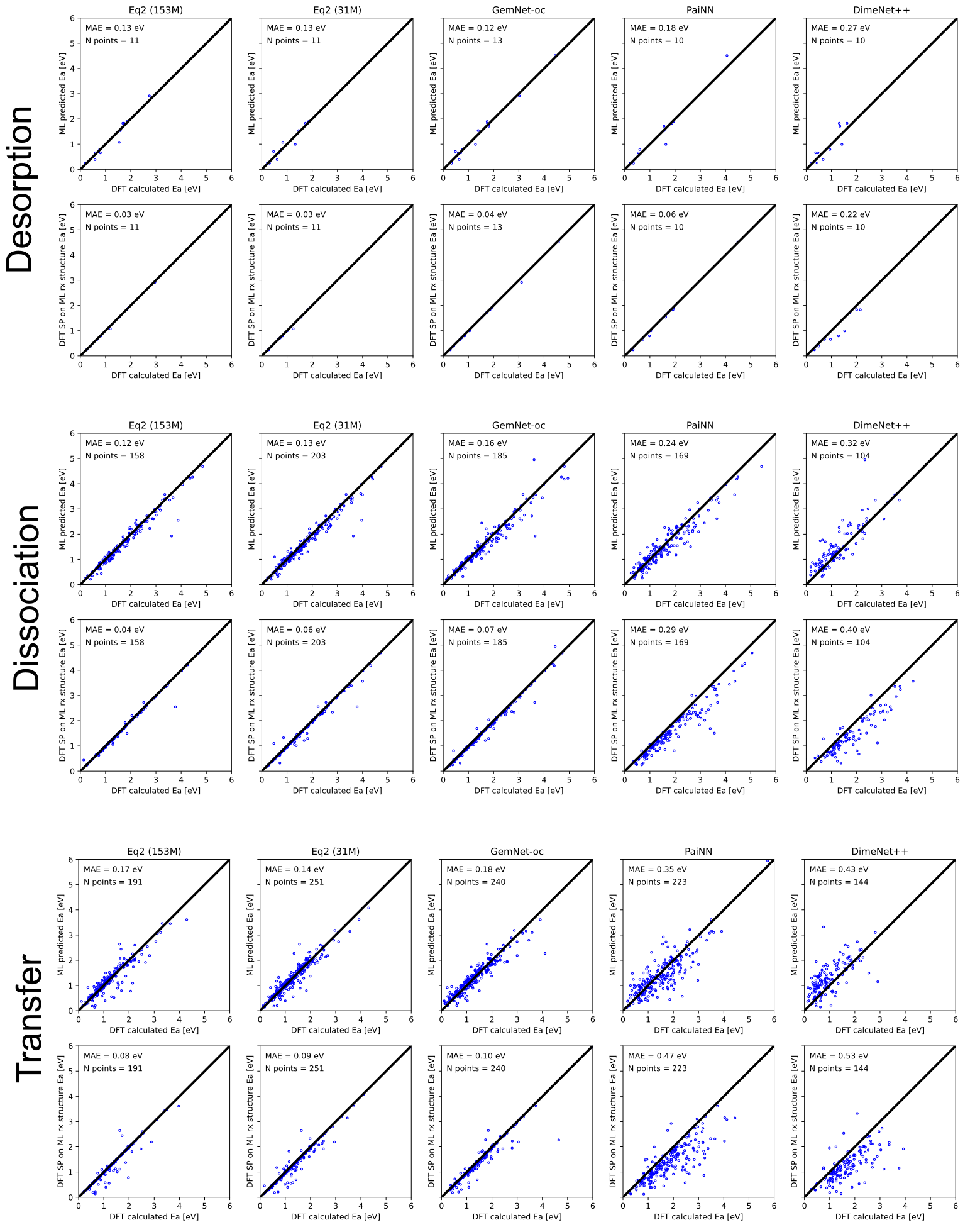}
    \caption{Parity plots with MAE for all models across the three reaction types}
    \label{fig:parity}

\end{figure}
\clearpage
\clearpage
\clearpage

\end{document}